\documentclass[twocolumn,aps,prb,reprint,groupedaddress]{revtex4}
\usepackage{amsmath}
\usepackage{amssymb}
\usepackage{etoolbox}
\usepackage{graphicx}
\usepackage{color}
\usepackage{booktabs}

\newcommand{\be}{\begin{equation}}
\newcommand{\ee}{\end{equation}}
\def\ba{\begin{aligned}}
\def\ea{\end{aligned}}

\newcommand{\bea}{\begin{eqnarray}}
\newcommand{\eea}{\end{eqnarray}}

\renewcommand{\Re}{{\rm \, Re\,}}
\renewcommand{\Im}{{\rm \, Im\,}}

\renewcommand{\hat}[1]{{\widehat #1}}

\renewcommand{\Im}{{\rm Im\,}}

\begin{document}

\title{
Multifractal states in  self-consistent theory of localization: analytical solution.}
\author{ B.~L.~Altshuler }
\affiliation{Physics Department, Columbia University, 538 West 120th Street, New York, New York 10027, USA }

 \author{ L.~B.~Ioffe }
\affiliation{CNRS and Universite Paris Sud, UMR 8626, LPTMS, Orsay Cedex, F-91405, France}
\affiliation{ L. D. Landau Institute for Theoretical Physics, Chernogolovka, 142432, Moscow region, Russia}

\author{V.~E.~Kravtsov
}
\affiliation{Abdus Salam International Center for Theoretical Physics, Strada Costiera 11, 34151 Trieste, Italy}
\affiliation{ L. D. Landau Institute for Theoretical Physics, Chernogolovka, 142432, Moscow region, Russia}

 \begin{abstract}
We consider disordered tight-binding models
which Green's functions obey the self-consistent {\it cavity equations}. Based on these equations and the replica representation, we derive an analytical expression for the  fractal dimension $D_{1}$ that distinguishes between
the extended ergodic, $D_{1}=1$, and extended non-ergodic (multifractal), $0<D_{1}<1$ states. The latter corresponds to the solution with broken replica symmetry, while the former corresponds to the replica-symmetric solution. We prove the existence of the extended non-ergodic phase in a broad range of disorder strength and energy as well as existence of transition between the two extended phases.  The results are applied to the systems with local tree structure (Bethe lattices) and to the systems with infinite connectivity (Rosenzweig-Poter random matrix theory). We obtain the phase diagram in the disorder-energy plain for the Bethe lattice and identify two insulating phases classified by the (one-step) replica symmetry breaking parameter.

\end{abstract}

\maketitle
\section{Introduction}
Recent progress in understanding the dynamical processes
of mesoscopic and macroscopic isolated disordered quantum Many-Body systems is
based on the concept of Many-Body Localization (MBL) \cite{BAA,Ogan-Huse,Ogan-Huse1}: the
Many-Body eigenstates can be localized in the Hilbert space in a way
similar to the conventional real space Anderson Localization \cite{Anderson} of a
single quantum particle by a quenched disorder. Depending on the temperature
(total energy) or other tunable parameters the system can find itself
either in MBL or in the Many-Body extended (MBE) phase. In the former
case the system of interacting quantum particles/spins cannot be described
in terms of conventional Statistical Mechanics: the notion of the
thermal equilibrium loses its meaning
There are serious reasons to believe that the violation
of the conventional thermodynamics does not disappear with the Anderson
transition (AT) from MBL to MBE state \cite{LevPino}: in a finite range
of the tunable parameters we expect the non-ergodic MBE phase (NMBE) where
the conventional theory is inapplicable although the many-body quantum
states are extended.

It is widely believed that in the one-particle Anderson problem in
the finite-dimensional space there is nothing like non-ergodic phase
-- all extended quantum states are ergodic and the ergodicity is violated
only at the critical point of the AT, which is manifested by the multifractality
of the critical quantum states \cite{Mirlin-rev}. This is likely to be due to the relatively
slow (polynomial) increase of the number of the quantum states with
the volume. Contrarily in the MBL problem the number of the original
states connected with a given one in the $n$-th order of the perturbation
theory in the interaction increases exponentially (or even quicker) with $n$ \cite{AGKL}.
Accordingly the one-particle problem on hierarchical lattices such
as the Bethe lattice (BL) (where the number of sites at a given distance
also increases exponentially) are likely to exhibit the main generic
features of the NMBE phases.

Recent numerical studies \cite{Biroli,Our-BL,Arx} of the Anderson problem on a random regular
graph (RRG), which is known to be almost indistinguishable from the
Bethe lattice at short length scales, brought up strong evidence in
favor of the existence of the non-ergodic phase: the eigenfunctions
were found to be multifractal with the fractal dimensions depending on disorder.
It was also possible to suggest that the transition (referred to below as {\it ergodic transition} (ET))
from the extended ergodic (EE) to the extended non-ergodic (NE) phases  is a true (first order) transition rather
than a crossover \cite{Arx}.

Existence of NE phase and the (second order) transition from NE to EE states
has been recently proven \cite{Rosen, RP-Bir} for an apparently different model: the random matrix theory
with the special diagonal suggested in 1960 by Rosenzweig and Porter (RP RMT) \cite{RPort}.
The concept that unifies both models is the self-consistent equations for the Green's function
suggested for the Bethe lattice by Abou-Chakra, Thouless and Anderson \cite{AbouChacAnd}. These equation are valid for the Bethe lattice
with any connectivity $K$ due to the loop-less, tree structure of BL. However, being a kind of self-consistent theory, these equations are
also valid exactly for the RP RMT, due to its infinite connectivity in the thermodynamic limit.

In this paper we develop an
analytical approach to the non-ergodic phase of the Anderson model
on the large-K Bethe Lattice.
We demonstrate that this approach can
be extended to a wide class of models including e.g. the  Rosentsweig-Porter
 model. In general our approach is applicable
for the models in which the loops either do not exist like on the
Bethe Lattice or are very rare as on RRG or are irrelevant as for
PR model.
\section{The models}
We are considering the Anderson model  on the graph with \textit{$N\gg1$}
sites:

\begin{equation}
\hat{H}=\sum_{i}^{N}\varepsilon_{i}\left\vert i\right\rangle \left\langle i\right\vert +\sum_{i,j=1}^{N}t_{ij}\left\vert i\right\rangle \left\langle j\right\vert \label{eq:H}
\end{equation}
Here \textit{$i=1,2,...,N$ } labels sites of the graph and \textit{$t_{ij}$
}is connectivity matrix of this graph: \textit{$t_{ij}$} equals to
$1$ if the sites \textit{$i$} and \textit{$j$} are connected,$\ $otherwise
\textit{$t_{ij}=0$}. This class of models is characterized by the
onsite disorder: $\epsilon_{i}$ are random on-site energies uniformly
distributed in the interval \textit{$(-W/2,W/2)$. }For the RRG problem
each site has $\mathit{K+1}$ neighbours while for infinite BL each
site is connected to \textit{K }neighbors of the previous generation
and one site in the next generation.

In the case of RP model the graph
is fully connected: each site is connected with everyone which is formally equivalent to $K=N$
In addition, in order to compensate for a macroscopic connectivity, the diagonal disorder strength
is proportional to the certain power $\gamma$ of $N$: $W\propto N^{\gamma/2}$, while the variance ${\rm var}\;t_{ij}=1$ (with the average $t_{ij}$ equal to 0). The Anderson transition in RP model
corresponds to $\gamma=2$, while the ergodic transition happens at $\gamma=1$  with the NE phase existing
in the interval $1<\gamma<2$ (see Ref.\cite{Rosen} and references therein).

\section{Distribution of LDoS and the  definition of spectral fractal dimension $D$.}
\begin{figure}[h]
\center{
\includegraphics[width=0.9\linewidth,angle=0]{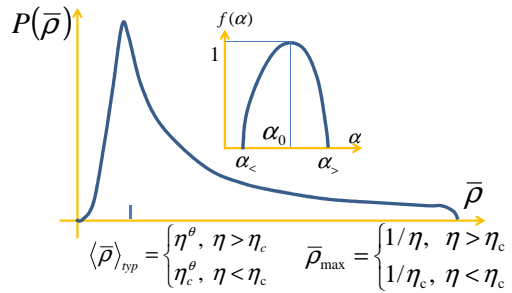}}
\caption{(Color online) Typical PDF $P(\bar{\rho})=\rho/\langle\rho\rangle$ for the multifractal states. {\it By definition} the typical average $\langle  \bar{\rho} \rangle_{{\rm typ}}\propto N^{-1+D}$ in the limit $\eta\ll \eta_{c}\propto N^{-z}$. In the inset: typical $f(\alpha)$ in the non-ergodic phase.}
\label{Fig:PDF}
\end{figure}
According to Ref.\cite{AbouChacAnd} the probability distribution function (PDF) $P(\bar{\rho}=\rho/\langle\rho\rangle)$ of the local density of states (LDoS) $\rho_{i}=(1/\pi)\,\Im G_{i}$:
\be\label{LDoS}
\rho_{i}(\omega)=\frac{1}{\pi}\sum_{a}|\langle i|a\rangle|^{2}\,\frac{\eta}{(\omega-E_{a})^{2}+\eta^{2}},
\ee
bears information on the character of wave functions $\psi_{a}(i)=\langle i|a\rangle $. Here  $G_{i}$ is the Green's function, $\langle\rho\rangle$ is the ensemble average LDoS, and $\eta$ is the broadening of energy levels.

For localized wave functions $P(\bar{\rho})$ is singular: for all $\bar{\rho} >0$ it vanishes
in the limit $\eta\rightarrow 0$, while for extended ergodic wave functions the limit $\eta\rightarrow 0$   leads to a stable non-singular $P(\bar{\rho})$. In both cases the limit $\eta\rightarrow 0$ is supposed to be taken {\it after} the limit $N\rightarrow\infty$.

Taking the limit $N\rightarrow\infty$ first eliminates the possibility to distinguish between the localized and the non-ergodic extended (multifractal) states. For such states the
typical wave function occupies a number of sites $M\propto N^{D}$ ($0<D<1$). Thus $M$ tends to infinity in the thermodynamic limit (extended wave function), yet it corresponds to a zero fraction of all the sites $N$ (hence non-ergodicity). In this case the PDF $P(\bar{\rho})$ shares both the property of the localized case and that of the ergodic extended case. Like in the localized case it develops a long tail at large $\bar{\rho}$ which spreads up to $\bar{\rho}_{max}=1/\eta$ with the typical $\bar{\rho}_{typ}\propto \eta^{\theta}$ ($\theta>0$) shrinking as $\eta$ decreases (see Fig.\ref{Fig:PDF}). However, like in the ergodic case this $\eta$-dependence saturates at $ \eta\ll \eta_{c}(N)\propto N^{-z}$ ($0<z<1$) resulting in $P(\bar{\rho})$ which is non-singular at any finite $N$. Existence of an energy scale $\eta_{c}(N)$ which decreases with $N$ but is much larger   than the mean level spacing $\delta\propto N^{-1}$, is the hallmark of the non-ergodic extended state. An example of such a behavior of $P(\bar{\rho})$ which can be computed analytically is presented in Ref.\cite{RP-Bir} for the RP model (where $\theta=1$ and $z=\gamma-1$) in the region of parameters $1<\gamma<2$ where the multifractality of wave functions has been earlier proven \cite{Rosen}.

Motivated by this picture we define the {\it spectral fractal dimension} $D$ in terms of the $N$-dependence of the typical average of $\Im G $:
\be\label{definition-D}
\frac{\langle \Im G \rangle_{\rm typ}}{\langle \Im G \rangle}=\langle \bar{\rho} \rangle_{\rm typ}=\eta_{c}^{\theta}\propto N^{-1+D},\;\;\;\;\eta\ll\eta_{c}\propto N^{-z}.
\ee
The exponent $D$ determines, via the scaling relationship:
\be\label{scaling}
z\theta=1-D,
\ee
the characteristic  energy scale $\eta_{c}(N)$ in the singular-continuous LDoS spectrum. Hence it is referred to as the {\it spectral fractal dimension}.

According to the definition Eq.(\ref{definition-D}) in the ergodic phase where the tail in Fig.\ref{Fig:PDF} collapses ($\langle \bar{\rho} \rangle_{\rm typ}\sim N^{-1+D}\sim 1/\eta_{c}\sim 1$) one obtains $D=1$, $z=0$. In the vicinity of the Anderson transition point where $\langle \bar{\rho} \rangle_{\rm typ}\sim \eta$ and $\eta_{c}\sim \delta \sim N^{-1}$ one obtains $\theta=1$, $z=1$ and hence $D=0$. Any value of $D$ in the interval $0<D<1$ corresponds to the non-ergodic extended (multifractal) phase.

\section{Abou-Chakra-Thouless-Anderson equations}

For a general lattice one can write self-consistent equations for
the \emph{two point} Green functions $G_{ij}=\left\langle i\right\vert \left(\omega-H\right)^{-1}\left\vert j\right\rangle $.
In the absence of the loops, it is possible to derive self-consistent
equations for the \emph{single site} Green functions, $G_{i}\equiv G_{ii}$
and $G_{i\rightarrow j}$ where the latter denotes single site Green
function with the bond $i\rightarrow j$ removed:
\begin{align}
G_{i\rightarrow k}  & =\frac{1}{\omega-\epsilon_{i}-\sum_{j\neq k}G_{j\rightarrow i} }\label{eq:Recursion}\\
G_{i}  & =\frac{1}{\omega-\epsilon_{i}-\sum_{j}G_{j\rightarrow i} }
\end{align}

For the Bethe lattice one can introduce the notion of generations:
each site of a given generation, $g$, is connected to $K$ ancestors
(generation $g+1$) and $1$ descendent (generation $g-1$) and focus
only on the Green functions $G_{i\rightarrow k}$ in which $k$ is
descendent of $i$: $G_{i\rightarrow m}=G^{(g)}_{i} $ where
\begin{equation}
G^{(g)}_{i}(\omega) =\frac{1}{\omega-\epsilon_{i}-\sum_{j(i)}G^{(g+1)}_{j}(\omega)}\label{eq:G_i^g(omega)}
\end{equation}
where $j(i)$ are ancestors of $i$.

Note that Eq.(\ref{eq:G_i^g(omega)}) should be also valid for the infinite number of
terms in the sum in the r.h.s. The reason is that it is essentially the mean-field equation
which is always exact in the limit of infinite connectivity $K$.

One can use the recursive relation
Eq.(\ref{eq:G_i^g(omega)}) to find the stationary distribution of
$G$. This approach was first employed by the authors of the seminal
paper \cite{AbouChacAnd} who used it to prove the existence of
the localized phase on Bethe lattice and to determine the critical disorder
$W_{c}$ of the AT. Recently we have generalized it to identify the non-ergodic
phase on Bethe lattice \cite{Arx}.

The equations Eq.\eqref{eq:G_i^g(omega)} are underdetermined: the
pole-like singularities in the right hand side of this equation have
to be regularized  by adding an infinitesimal
imaginary part $\eta$ to $\omega\rightarrow\omega+i\eta$ similar to Eq.(\ref{LDoS}).
The recursion Eq.\eqref{eq:G_i^g(omega)} might become unstable with
respect to this addition.
This instability indeed occurs  for $W < W_{c}$ and marks the onset of the extended phase, while for $W>W_{c}$ the solution with
$P(G)\propto\delta(\Im G)$ is stable.

\section{Increment $\Lambda$ of $\Im G$ and spectral fractal dimension $D$}
The spectral fractal dimension $D$ can be obtained from the modified recursion equation:
\begin{equation}
G_{i}^{(g)}(\omega,\eta)=\frac{1}{\omega-\epsilon_{i}+i\eta-\sum_{j(i)}G_{j}^{(g+1)}(\omega,\eta)}.\label{eq:G_open}
\end{equation}
The  recursive  procedure Eq.(\ref{eq:G_open}) is the basis for the recursive algorithm known as {\it population dynamics} (PD) \cite{pop-dyn} which corresponds to an infinite tree $N\rightarrow \infty$.
At an infinitesimal $\eta>0$ and $W<W_{c}$ the typical imaginary part $\langle \Im G\rangle_{typ}$ increases exponentially with the number of recursion steps $\ell$ in Eq.(\ref{eq:G_open}):
\be\label{Lambda}
\langle\Im G\rangle_{typ} \propto\eta\, e^{\Lambda\, \ell},
\ee
where $\Lambda$ is the corresponding  increment.

In order to relate the results of PD  with any finite system one has to terminate the exponential growth by a sort of physical argument which lies {\it outside}
of the recursive procedure. The simplest  case is the problem of eigenfunction statistics in the root of a finite BL.   In this case the maximal number of generations $g_{{\rm max}}=\ln N/\ln K$ which naturally terminates the recursive procedure at $\ell=\ell_{t}=\ln N/\ln K$.  For the case of RRG it is natural to assume that the termination of exponential growth Eq.(\ref{Lambda}) happens when a first loop will be almost surely completed \cite{almost-surely}.  For RRG this implies $\ell_{t}=R$, where $R=\ln N/\ln K + O(\ln \ln N)$ \cite{Bollobas} is the diameter of the graph.

In addition to termination of the recursive procedure a finite $N$ sets the lower bound $\eta>\delta\sim N^{-1}$ for the level broadening which is necessary for distinguishing between the localized and extended phases.
Thus the limit $\eta\rightarrow 0$ for finite systems implies $\eta\sim \delta$ in Eq.(\ref{Lambda}).

With this finite-$N$ modifications the expression for $\langle \Im G\rangle_{typ}$ takes the form:
\be\label{ImG-Lambda}
\langle\Im G\rangle_{typ}\left|_{\eta\rightarrow 0}\right. \propto  N^{-1+\Lambda/\ln K}.
\ee
Comparing Eq.(\ref{ImG-Lambda}) with  the definition Eq.(\ref{definition-D})   of the spectral fractal dimension $D$, we arrive at:
\be\label{Lambda-D}
D=\frac{\Lambda}{\ln K}.
\ee
We will see below that Eq.(\ref{Lambda-D}) applies also to the RP RMT where each site is connected with any other site and $K=N$. This corresponds to termination of the recursion just after one step
$\ell_{t}=1$, since the next step will inevitably create a loop.

We conclude, therefore, that Eq.(\ref{Lambda-D}) is a basic consequence of the Abou-Chakra-Thouless-Anderson theory which is valid for a broad class of hierarchical as well as mean-field-like systems.

\section{Spectral fractal dimension $D$ and the information fractal dimension $D_{1}$}
A broad class of measures similar to Fig.\ref{Fig:PDF} is given for $\eta\ll\eta_{c}$ by the {\it multifractal ansatz} \cite{Our-BL}:
\be\label{multifractal-ansatz}
P(\bar{\rho})d\bar{\rho} = \frac{d\bar{\rho}}{\bar{\rho}}\,A\, N^{f(\alpha)-1},\;\;\;\;\alpha=1-\frac{\ln \bar{\rho}}{\ln N}.
\ee
The function $f(\alpha)$ is defined for $\alpha\geq 1-z$
(as $\bar{\rho}_{max}\propto N^{z}$), where
\be\label{d-sc}
0\leq z\leq 1-\alpha_{<}\,\leq 1,
\ee
and should have a maximum at some point $\alpha=\alpha_{0}$ such that
$f(\alpha_{0})=1$. This ensures the normalization condition $\int P(\bar{\rho}) d\bar{\rho} =1$ at the normalization constant $A$ which is only slowly (logarithmically)
$N$-dependent. The exponent $\alpha_{0}$ determines the most abundant (typical) value of $\bar{\rho}$:
\be\label{g-typ}
\langle \bar{\rho}\rangle_{\rm typ}={\rm exp}\left[ \int \ln \bar{\rho}\, P(\bar{\rho})\,d\bar{\rho}\right] \sim N^{1-\alpha_{0}}.
\ee
Comparing Eqs.(\ref{definition-D}),(\ref{g-typ}) we immediately arrive at:
\be\label{D-alpha-0}
D=2-\alpha_{0}.
\ee
It is absolutely natural that the spectral fractal dimension defined in terms of the typical imaginary part of Green's function is expressed
in terms of the exponent $\alpha_{0}$ which is related with the most abundant value of LDoS.

Now we relate the spectral fractal dimension $D$ with the {\it information fractal dimension} $D_{1}$
defined \cite{Mirlin-rev} through the averaged Shannon entropy of a random wave function $\psi$:
\be
\ln \langle S \rangle = \left\langle  \sum_{i} |\psi(i)|^{2}\,\ln(|\psi(i)|^{2})\right\rangle = -D_{1}\,\ln N + {\rm const}.
\ee
The quantity $D_{1}$ is also equal to the Hausdorff fractal  dimension of the support set of a multifractal wave function \cite{Our-sset}.

In order to relate $D$ and $D_{1}$ we employ the duality of $P(\bar{\rho})$:
\be\label{dual}
P(1/\bar{\rho})= \bar{\rho}^{3}\,P( \bar{\rho}),
\ee
which implies a constraint:
\be\label{MF-sym}
f( \alpha)=f(2-\alpha)+\alpha-1.
\ee
for  $f(\alpha)$ in  Eq.(\ref{multifractal-ansatz}). The duality Eq.(\ref{dual}) was first discovered in Ref.\cite{Alt-Prig} for the LDoS distribution in strictly one-dimensional systems using the Berezinskii diagrammatics \cite{Ber} and later on derived for systems of {\it any} dimensions in the framework of the nonlinear supersymmetric sigma model \cite{MF}. This duality is valid under very broad conditions \cite{Savin-Fyod} in both localized and extended phases.
As $f(\alpha_{0})=1$ one immediately finds from Eq.(\ref{MF-sym}):
\be\label{smal1}
D=2-\alpha_{0}=f(2-\alpha_{0}).
\ee
On the other hand, $\alpha_{q}$ defined as the root of $f'(\alpha_{q})=q$, obeys at $q=1$ the equation
$$f(\alpha_{1})-\alpha_{1}=0,$$
which follows from the normalization condition $\langle \bar{\rho}\rangle=\int \bar{\rho}\,P(\bar{\rho})\,d\bar{\rho}=A\ln N\int_{0}^{\infty}N^{f(\alpha)-\alpha}\,d\alpha=1$. Then uniqueness of the solution
allows to conclude that:
\be\label{D-D1}
D=\alpha_{1}=D_{1}.
\ee
The latter equality follows from the fact that
\bea
D_{1}&=& \lim_{q\rightarrow 1} D_{q}\equiv\frac{q\alpha_{q}-f(\alpha_{q})}{q-1}=\nonumber \\ &=&\alpha_{1}+\frac{\partial \alpha_{q}}{\partial q}\,\left(q-f'(\alpha_{q}\right))\left|_{q=1} \right..
\eea

Thus we conclude that the spectral fractal dimension $D$ defined in Eq.(\ref{definition-D}) is nothing but the information fractal dimension $D_{1}$.

\section{Large connectivity approximation}
The increment $\Lambda$  for a small connectivity $K=2$ was computed  numerically  in Ref.\cite{Arx} using the PD algorithm \cite{pop-dyn} based on Eq.(\ref{eq:G_open}). It is a continuous function of disorder $W$ that vanishes at the AT point $W=W_{c}$ and grows as $W$ decreases below $W_{c}$ in an almost linear manner with $D$ being smaller than 1 for all $W > 7.5$ where  a satisfactory convergence of PD was reached. In this section we derive an analytic expression for $\Lambda(W)$ and $D_{1}(W)$ in the case of a large connectivity $\ln K\gg 1$.

Linearizing the r.h.s. of Eq.(\ref{eq:G_open}) in $\Im G$ we obtain:
\begin{equation}
\Im G_{i}^{(g)}(\omega)=\frac{\sum_{j(i)}\Im G_{j}^{(g+1)}}{\left(\omega-\epsilon_{i}-\Re \Sigma_{i}^{(g+1)}\right)^{2}}\label{eq:ImG_lin}
\end{equation}
The general method for the solution of such equations was developed
in \cite{Derrida} that used mapping to traveling wave problem.
More compact solution uses replica approach and one step replica symmetry
breaking (see e.g. \cite{IoffeMezard}).

 We begin with the expression $\Lambda(\omega)=\overline{\ln Z(\omega)}/\ell$ for $\omega$- (and $W$-) dependent
 $\Lambda(\omega)$,
where
\begin{equation}
Z(\omega)=\sum_{P}\prod_{k=1}^{\ell}\frac{1}{\left(\omega-\epsilon^{(k)}_{P}\right)^{2}}.\label{eq:f_i}
\end{equation}
In Eq.(\ref{eq:f_i}) $P$ determines a path that goes from
the initial point to a point in the generation $\ell$ ($k=1...\ell$).

The function
$f$ has a meaning of the free energy of a polymer on the Bethe lattice
\cite{Derrida}with unusual disordered site energies $\beta E_{j}=\ln\left(\omega-\epsilon_{j}\right)^{2}$.

In order to compute the free energy Eq.(\ref{eq:f_i}) we use replica
method
\begin{eqnarray*}
\Lambda(\omega) & =\lim_{n\rightarrow0} & \frac{1}{n\ell}\left(\overline{Z^{n}}-1\right).
\end{eqnarray*}
Where $\overline{Z^{n}}$ can be written as:
\be\label{all-pathes}
\overline{Z^{n}}=\overline{\sum_{P_{1},...P_{n}}\prod_{a=1}^{n}\prod_{k=1}^{\ell} \frac{1}{(\omega-\varepsilon^{(k)}_{P_{a}})^{2}}}.
\ee
Without replica symmetry breaking (RSB) there would be $K^{\ell n}$ different
pathes contributing to the free energy $Z(\omega)$. The stable solution
corresponds to one step replica symmetry breaking in which $n$ pathes
are grouped into $n/m$ groups of $m$ {\it identical} pathes each, considering the contribution of different groups as statistically independent.
The RSB solution for the increment $\Lambda$ is then obtained by minimization of so obtained  $\Lambda(\omega,m)$ with respect to $m$:
\be\label{min}
\Lambda(\omega)={\rm min}_{m}\Lambda(\omega,m)\equiv \Lambda(\omega,m_{0}).
\ee
Despite  an apparent neglect of more complicated correlations between pathes, such an approach appears to be exact in many cases (see e.g. \cite{IoffeMezard}).

The next step is averaging w.r.t. random on-site energies:
\begin{eqnarray}
\Lambda(\omega,m) & =&\lim_{n\rightarrow0}\frac{1}{n}\left[\left(K\int F(\epsilon)\,\frac{d\epsilon}{\left|\omega-\epsilon\right|^{2m}}\right)^{n/m}-1\right] \nonumber \\
 & =&\frac{1}{m}\ln\left(K\int F(\epsilon )\, \frac{d\epsilon}{\left|\omega-\epsilon\right|^{2m}}\right),
 \label{Lambda-averaging}
\end{eqnarray}
where $F(\epsilon)=(1/W)\,\theta(W/2 -|\epsilon|)$.

It is this step where we employ the large $K$ approximation.
As it has been shown in Ref.\cite{AbouChacAnd}, under the assumption of large disorder $W$ which is relevant for large connectivity $K$,  the real part of the self-energy in the denominator of Eq.(\ref{eq:ImG_lin}) can be neglected.
Then the increment $\Lambda(\omega)$ found from Eq.(\ref{Lambda-averaging}) takes the form:
\be\label{Lambda-fin-large-K}
\Lambda=2\,\ln \left(\frac{W_{c}(\omega)}{W}\right).
\ee

The critical disorder $W_{c}=W_{c}(0)$ in Eq.(\ref{Lambda-fin-large-K})
 close to the middle of the band  is defined as:
\be
\ln\frac{W_{c}}{2}= \frac{1}{2m_{0}}\ln\frac{K}{1-2m_{0}}.\label{eq:lnWc/2}
\ee
where $m_{0}$ is found from the minimization condition Eq.(\ref{min}):
\be \label{mstar}
\frac{2m_{0}}{1-2m_{0}}= \ln\frac{K}{1-2m_{0}}
\ee
Combining (\ref{eq:lnWc/2},\ref{mstar}) to exclude $1/(1-2m_{0})$
we get an equation for $W_{c}$:
\be\label{upper-limit}
K\,\ln\left(\frac{W_{c}}{2}\right)=\frac{W_{c}}{2e}.
\ee
 At large $K\gg1$ one obtains with logarithmic accuracy
\be\label{lKWc}
W_{c}\approx 2eK\ln (eK)
\ee
in agreement with \cite{AbouChacAnd}.
\section{RSB parameter $m_{0}$ and Abou-Chakra-Thouless-Anderson exponent $\beta$}
As a matter of fact
$m_{0}$ found from Eq.(\ref{mstar}) (which in the above approximation is independent of $W$) has a special physical meaning. Namely, it is related with the power of $\bar{\rho}$ (or $N |\psi|^{2}$, see Appendix A) in the power-law
distribution function $P(\bar{\rho})$ (or $P(N |\psi|^{2})$) at $W=W_{c}$:
\be\label{PDF}
P(\bar{\rho})\sim \frac{\langle \bar{\rho}\rangle_{typ}}{(\bar{\rho})^{1+m_{0}}},
\ee
This fact can be confirmed (see Appendix A) by comparing
the equation for this power (at $W=W_{c}$) derived in Ref.\cite{Our-BL} and Eq.(\ref{mstar}).

Thus the RSB parameter $m_{0}$ at $W=W_{c}$ is {\it identical} to the exponent $\beta$ introduced in Ref.\cite{AbouChacAnd}. Note that this exponent was shown in Ref.\cite{AbouChacAnd} to be equal to:
\be\label{beta-c}
\beta=1/2
\ee
and Eq.(\ref{beta-c}) was used as a  condition to compute $W_{c}$. The same result, Eq.(\ref{beta-c}), follows from the
duality Eqs.(\ref{dual}),(\ref{MF-sym}) for a linear $f(\alpha)$.

On the other hand,  one obtains from Eq.(\ref{mstar}):
\be\label{m_0}
m_{0}\approx 1/2-1/(2\ln K).
\ee
This means that the exact duality is violated  and the accuracy of the large-K approximation is $O(1/\ln K)$.

\section{Minimal account for the real part of self-energy}
 The fact that Eq.(\ref{m_0}) gives $1/\ln K$ corrections to $m_{0}=1/2$  is related with the
neglect of the real part of self-energy in the denominator of Eq.(\ref{eq:ImG_lin}) and the resulting logarithmic divergence of the average  $\int^{W/2}_{-W/2}\frac{d\epsilon}{W}\,\epsilon^{-2m}$ at $m=1/2$. Let us try to improve our derivation
in such a way that Eq.(\ref{beta-c}) is respected. In order to reach this goal we introduce in Eq.(\ref{Lambda-averaging}) the
{\it effective} distribution function $F_{\rm eff}(\epsilon)$  of the real part of  $\epsilon_{i}+\sum_{j(i)} G_{i}=\omega-G_{i}^{-1}$ instead of the
distribution
$F(\epsilon)$ of the on-site
energies $\epsilon_{i}$.  To avoid confusion we emphasize that this is an {\it effective distribution} which takes into account correlations of different
$[\Re G_{i}]^{2}$
in the product in Eq.(\ref{eq:f_i}) when $\Re \Sigma$ is not neglected. Indeed, if one of $\Re G_{j(i)}$ is anomalously large, the neighboring $\Re G_{i}\sim [\Re G_{j(i)}]^{-1}$
should be anomalously small. It is shown in Appendix B that this effect leads to the symmetry ${\cal P}(y)={\cal P}(1/y)$ of the PDF of
the product $y=\prod_{k}|G_{i_{k}}|^{-1}$ along a path $P$, which is equivalent to the corresponding symmetry of
$F_{\rm eff}(\epsilon)$:
\be\label{symmetry-omega}
F_{\rm eff}(\epsilon+\omega)=F_{\rm eff}(\epsilon^{-1}+\omega).
\ee
The simplest deformation the distribution function $F(\epsilon)$ in order to obey the symmetry Eq.(\ref{symmetry-omega}) at $\omega=0$ 
is the following:
\be\label{low-cutoff}
F_{\rm eff}(\epsilon)=F_{\rm eff}(1/\epsilon)=\frac{\theta(|\epsilon|-2/W)\theta(W/2 -|\epsilon|)}{W-4/W},
\ee
Thus the minimal account of $\Re \Sigma$ is equivalent to imposing  the symmetry Eq.(\ref{low-cutoff}) which eliminates small $|\epsilon|<2/W$. We will see
below that this is a crucial step, as it allows for the {\it replica-symmetric} solution which corresponds to $D=1$. With the previous distribution $F(\epsilon)$
this solution did not exist as $\int F(\epsilon)\,|\epsilon|^{-2m}\, d\epsilon$  diverges at $m=1$.

Now, the critical disorder $W_{c}$ and $m_{0}$ are found from the solution of the system of equations:
\be\label{system}
\Lambda(m,W)=0, \;\;\;\frac{\partial\Lambda}{\partial m}=\frac{2}{m}\frac{\int F_{\rm eff}(\epsilon)\,\ln(1/\epsilon)\,\frac{d\epsilon}{|\epsilon-\omega|^{2m}}}
{\int F_{\rm eff}(\epsilon)\,\frac{d\epsilon}{|\epsilon-\omega|^{2m}}}=0.
\ee
At $\omega=0$ replacing $\epsilon\rightarrow 1/\epsilon$ and using the symmetry of $F_{\rm eff}(\epsilon)$ one immediately sees that the integral 
in the numerator of $\frac{\partial\Lambda}{\partial m}$ changes sign and is thus equal to zero if $m=1/2$. We conclude 
that $m_{0}=1/2$ {\it exactly} at the AT point merely due to the symmetry Eq.(\ref{low-cutoff}).

The critical disorder $W_{c}$ is then found from the solution of the first of Eq.(\ref{system}) at $m=1/2$ using a concrete form, Eq.(\ref{low-cutoff}) 
of $F_{\rm eff}(\epsilon)$. The resulting equation for $W_{c}$ at $\omega=0$ is as simple as this:
\be\label{Wc-K}
2K\,\ln\left(\frac{W_{c}}{2}\right)=\frac{W_{c}}{2}-\frac{2}{W_{c}},
\ee
which should be compared with the "upper limit" Eq.(\ref{upper-limit}) due to Abou-Chacra, Thouless and Anderson.

The results of solution to this {\it algebraic} equation for different connectivity $K$ is summarized in  Table I.

 \begin{table}[h!]
  \centering
  \caption{Comparison of values for $W_{c}(K)$ obtained from Eq.(\ref{Wc-K}), from the "upper bound" of Ref.\cite{AbouChacAnd} (Eq.(\ref{upper-limit}) of this paper) and from numerics of Ref.\cite{Biroli-Tarzia-rev-BL}.}
  \label{tab:table1}
  \begin{tabular}{|c||c|c|c|c|c|c|c|}
   \hline
    K & 2 & 3 & 4 & 5 & 6 & 7 & 8\\
    \hline
    Eq.(\ref{Wc-K})  & 17.65 & 34.18 & 52.30 & 71.62 & 91.91 & 113.0 & 134.8 \\
    \hline
    Ref.\cite{Biroli-Tarzia-rev-BL} & 17.4 & 33.2 & 50.1 & 67.7 & 87.3 & 105 & 125.2 \\
    \hline
    Eq.(\ref{upper-limit}) & 29.1 & 53.6 & 80.3 & 108 & 138 & 169 & 200 \\
    \hline
  \end{tabular}
\end{table}
One can notice an {\it excellent} agreement with numerics even for the minimal $K=2$. The results for larger $K$ should be even more accurate.

We conclude that the correct symmetry improves at lot the large-K approximation and leads to an extremely simple and powerful formula for $W_{c}$
which accuracy exceeds by far any approximations to the exact Abou-Chacra-Thouless-Anderson theory known so far.
\section{Analytical results for $D(W)$ and $m(W)$ at the band center $\omega=0$.}
\begin{figure}[h]
\center{
\includegraphics[width=0.9\linewidth]{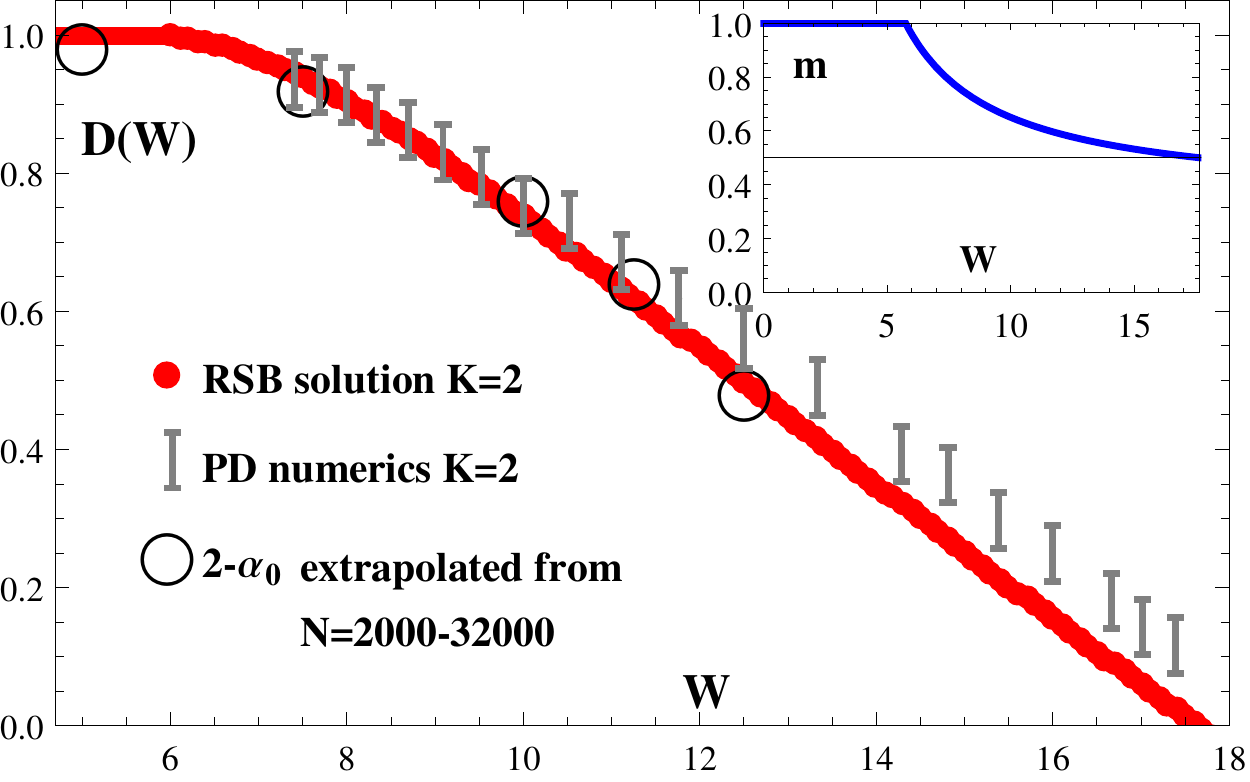}
}
\caption{(Color online) Fractal dimension $D(W) $ (red fat curve) found from the RSB solution, Eqs.(\ref{Lambda-D}),(\ref{min}),(\ref{Lambda-averaging}),(\ref{low-cutoff}), for $K=2$ compared with the results of the population dynamics numerics of Ref.\cite{Arx} (grey error bars) and $2-\alpha_{0}(W)$ (open circles) extracted from $f(\alpha)$ extrapolated from $N=2000-32000$ to $N=\infty$ using procedure and data of Ref.\cite{Our-BL}. The  non-ergodic extended phase which corresponds to RSB solution,    is stable for $W_{0}<W<W_{c}$, where $W_{0}\approx 5.74$ is the limit of stability of the non-ergodic phase and $W_{c}\approx 17.65$ is the AT point. Inset: dependence   on $W$ of $m_{0}$ minimizing $\Lambda(m)$. At $W=W_{c}$,  $m_{0}=1/2$ exactly, and at $W=W_{0}$ merging of RSB solution ($m_{0}\leq 1$)  with RS solution ($m=1$) occurs.
 }
\label{Fig:coincidence}
\end{figure}

Plugging Eq.(\ref{low-cutoff}) into Eq.(\ref{Lambda-averaging}) one can compute the fractal dimension $D(W)=D_{1}(W)$ for $K=2$ and compare it with
the numerical results of Refs.\cite{Arx, Our-BL} on RRG.
The results are summarized in Fig.\ref{Fig:coincidence}).
One can see an excellent agreement with the PD numerics of Ref.\cite{Arx} and a fairly good agreement with the $2-\alpha_{0}$ extracted from  $f(\alpha)$ extrapolated from relatively small sizes $N=2000-32000$ to $N=\infty$ using the numerics and procedure of Ref.\cite{Our-BL}.

In the inset of Fig.\ref{Fig:coincidence} the optimal $m=m_{0}$ minimizing $\Lambda(\omega,m)$ is shown. For the effective distribution $F_{\rm eff}(\epsilon)$ given by Eq.(\ref{low-cutoff}) we obtain $m_{0}=m(W)$ which is disorder-dependent. As  $W$ decreases below $W_{c}$, $m_{0}$ increases monotonically from $m_{0}=1/2$ and at some point $W=W_{0}$ it reaches $m_{0}=1$. At this point the RSB solution terminates, as the values of $m_{0}>1$ are unphysical  since the corresponding distribution function Eq.(\ref{PDF})  fails to simultaneously  fulfil two normalization conditions $\langle 1 \rangle=1$ and $\langle \bar{\rho}\rangle =1$.

This proves existence of the {\it ergodic transition} from the
non-ergodic extended (multifractal) phase described by the RSB solution to the extended ergodic phase described by the {\it replica symmetric} (RS) solution 
with $m=1$. Existence of such a RS solution and the fact that $D=1$ at $m=1$ is  a consequence of the symmetry Eq.(\ref{low-cutoff}). 
  Indeed, at $m=1$
(and $\omega=0$) we have:
\be\label{Lambda-RS}
\Lambda= \ln\left( K \int F_{\rm eff}(\epsilon)\,\frac{d\epsilon}{\epsilon^{2}} \right).
\ee
Because of the symmetry of $F_{\rm eff}(\epsilon)=F_{\rm eff}(1/\epsilon)$, changing the variables of integration $\epsilon\rightarrow 1/\epsilon$ converts 
the integral in Eq.(\ref{Lambda-RS}) into the normalization integral for the effective distribution function $\int F_{\rm eff}(\epsilon)\,d\epsilon =1$. Then we immediately obtain from Eq.(\ref{Lambda-D}) that the RS solution corresponds to $D=1$, i.e. to the ergodic extended phase.

\begin{figure}[h]
\center{
\includegraphics[width=0.9\linewidth]{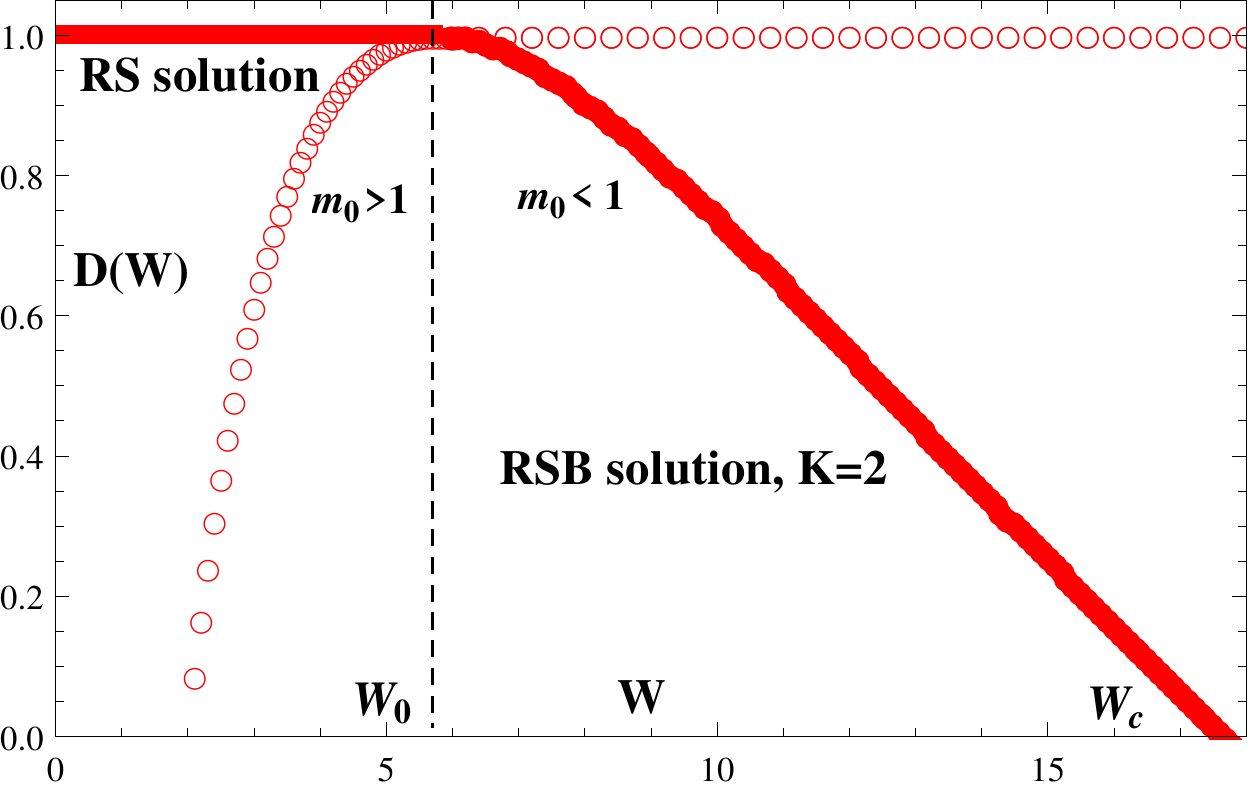}
}
\caption{(Color online) RSB and RS solutions to  Eqs. (\ref{Lambda-D}),(\ref{min}),(\ref{Lambda-averaging}),(\ref{low-cutoff}). 
The branch of the curve with $m_{0}>1$ (open circles) is unphysical, as the corresponding $P(\bar{\rho})$ is not normalizable. For $W<W_{0}$ only the RS solution with $D_{RS}=1$ is valid. For $W_{0}<W<W_{c}$ both solutions exist but only the one with the minimal $\Lambda$ is realized in PD calculations. }
\label{Fig:D_RSB}
\end{figure}
\begin{figure}[h]
\center{
\includegraphics[width=0.9\linewidth]{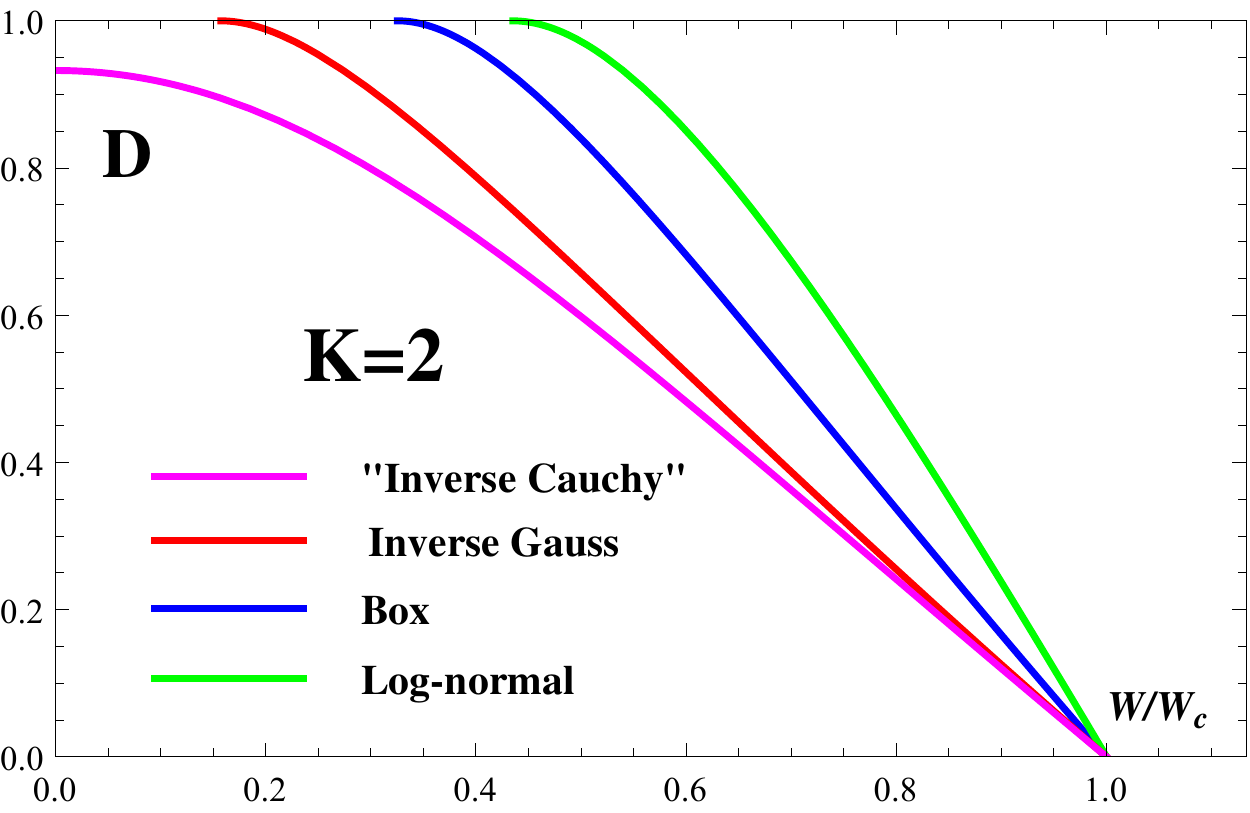}
}
\caption{(Color online) Spectral fractal dimension $D$ as a function of $W/W_{c}$ for the 
box-shaped $F_{\rm eff}(\epsilon)$ (Eq.(\ref{low-cutoff})); for the inverse Gauss $F_{\rm eff}(\epsilon)=C\,e^{\frac{2}{W}(\epsilon+\frac{1}{\epsilon})}$ , 
for the log-normal $F_{\rm eff}(\epsilon)=C\,e^{\ln^{2}(\epsilon)/\ln^{2}(W/2)}$, and for the "Inverse Cauchy" $F_{\rm eff}(\epsilon)=
f_{{\rm Cauchy}}(\epsilon+\epsilon^{-1})=C\,\frac{1}{\left((\epsilon+\epsilon^{-1})^{2}+(W/2)^{2}\right)}$. The termination point of the RSB solution 
at $W=W_{0}\neq 0$ corresponds to $D=D_{0}=1$ for all the cases except the Cauchy distribution where $W_{0}=0$ and $D_{0}<1$.  }
\label{Fig:differentF}
\end{figure}

Note that existence of termination point of the RSB solution at a non-zero $W_{0}$ is a generic feature of the theory. 
It occurs at any function $F_{\rm eff}(\epsilon)$ obeying the symmetry Eq.(\ref{low-cutoff}) and decreasing {\it sufficiently fast} 
at large and small $\epsilon$ (see Fig.\ref{Fig:differentF}). Only if $F_{\rm eff}(\epsilon)$ decays as $\epsilon^{\pm 2}$ (the "inverse Cauchy" distribution), the termination point is at $W_{0}=0$ and the corresponding $D<1$.

We would like to emphasize that $W_{0}$ is a {\it limit of stability} of the non-ergodic extended phase. The actual ergodic transition may occur {\it before} this limit is reached, as
the RS solution exists in the entire region $W<W_{c}$. In this  case it should be a first order transition at $W=W_{E}$ similar to the one observed in Ref.\cite{Arx}.
In Sec. XII we will formulate a plausible conjecture about the location of $W_{E}$.

\section{ Application to RP RMT}
\begin{figure}[h]
\center{
\includegraphics[width=0.9\linewidth]{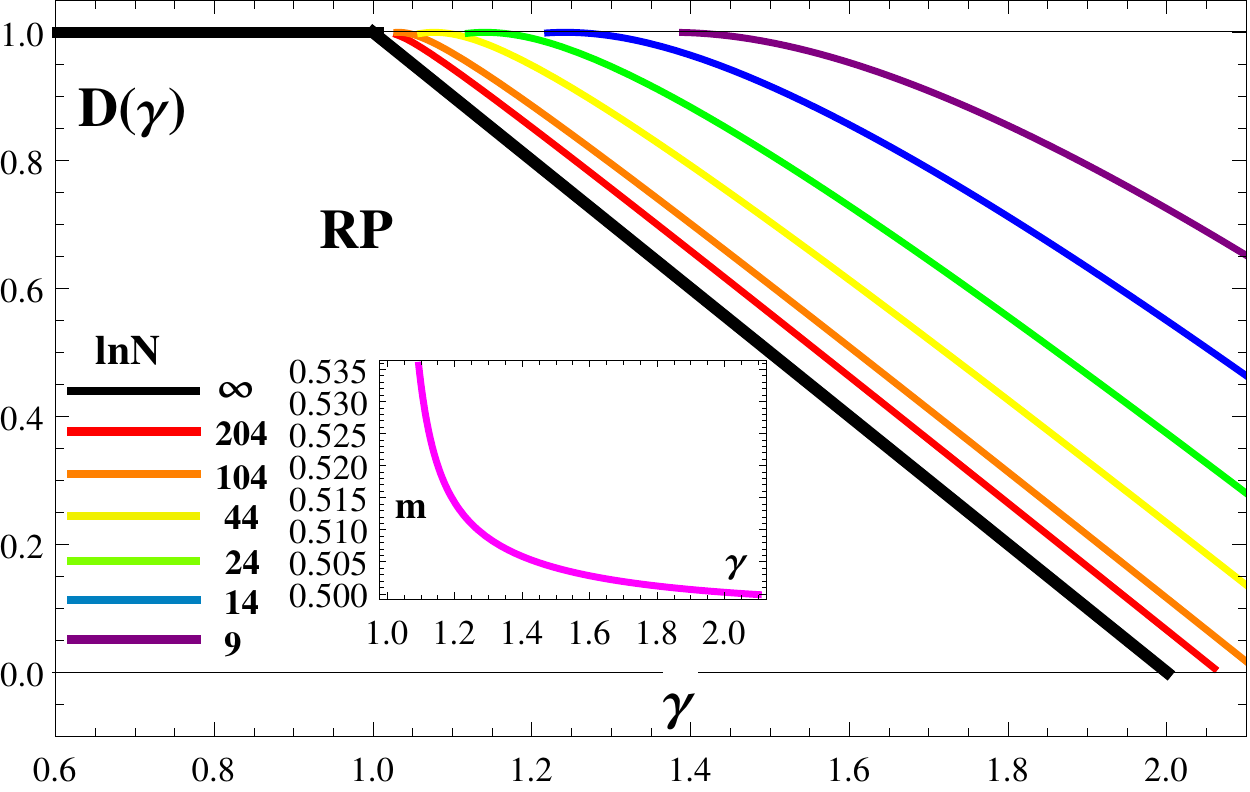}
}
\caption{(Color online)  Dependence  $D(\gamma)$ for the Rosenzweig-Porter RMT at different matrix sizes $N$ obtained from (\ref{Lambda-D}),(\ref{min}),(\ref{Lambda-averaging}),(\ref{low-cutoff}). In the limit $N\to\infty$, $D(\gamma)=2-\gamma$ coincides with the result of Ref.\cite{Rosen}. This confirms existence of non-ergodic extended phase in RP RMT predicted in Ref.\cite{Rosen} for $1<\gamma<2$. Inset: dependence on $\gamma$ of  $m_{0}$ minimizing $\Lambda(m)$ for $\ln N=204$. In the limit $N\to \infty$, $m_{0}=1/2$ all the way down to the limit of stability of RSB solution at $\gamma=1$, and then it jumps to $m=1$ in the ergodic extended phase at $\gamma<1$.}
\label{Fig:RP}
\end{figure}

The Rosenzweig-Porter random matrix theory is formally defined \cite{RPort,Rosen} as a Hermitean $N\times N$ matrix with random Gaussian entries $H_{ij}$ independently fluctuating about zero with the variance $\langle |H_{ii}|^{2}\rangle =1$, and $ \langle |H_{i\neq j}|^{2}\rangle = \lambda \,N^{-\gamma}$, where $\lambda$ is an $N$-independent number. By changing the energy scale one may define $h_{ij}$, where $\langle |h_{ii}|^{2}\rangle =\lambda^{-1}\,N^{\gamma}$ and $ \langle |h_{i\neq j}|^{2}\rangle =1$. Thus the RP model corresponds to $W\sim N^{\gamma/2}$. The AT critical point in the limit $N\to\infty$ corresponds to $\gamma=2$  (see Ref.\cite{Rosen} and references therein)  and thus $W_{c}\sim N$. Then Eqs.(\ref{Lambda-D}),(\ref{Lambda-fin-large-K}) result in:
\be\label{RP-D}
D(\gamma)=2-\gamma + O(1/\ln N),
\ee
which is valid exactly in the limit $\ln N\to\infty$ (see Fig.\ref{Fig:RP}).
This result coincides with the fractal dimensions $D_{q}=2-\gamma$ (valid for all $q>1/2$) for the RP RMT obtained in Ref.\cite{Rosen} by completely different arguments.   Note that $m_{0}$ minimizing $\Lambda(m)$ is 1/2 in the entire region of non-ergodic extended states $1<\gamma<2$ in the limit $\ln N\rightarrow\infty$.
This implies that the exponent in the
power-law PDF Eq.(\ref{PDF}) (as well as in the wave function amplitude distribution $P(\ln|\psi|^{2})\propto (1/|\psi^{2}|)^{m_{0}}$) is $1/2$ for all values of $\gamma>1$, i.e. $f(\alpha)=\frac{1}{2}\alpha +\frac{1}{2}(2-\gamma)$ ($2-\gamma<\alpha<\gamma$) in agreement with Ref.\cite{Rosen}. Because of the abrupt cutoff of $f(\alpha)$ at $\alpha=\alpha_{<}=2-\gamma$ \cite{Rosen} the dynamical scaling exponent, Eq.(\ref{d-sc}), $z=1-\alpha_{<}=1-D(\gamma)=\gamma-1$ \cite{Rosen,RP-Bir} and the exponent $\theta$, Eq.(\ref{scaling}), is equal to 1 \cite{RP-Bir}.
\section{Phase diagram and the Lyapunov exponents}
\begin{figure*}[th]
 \centering{
 \includegraphics[width=0.45\linewidth]{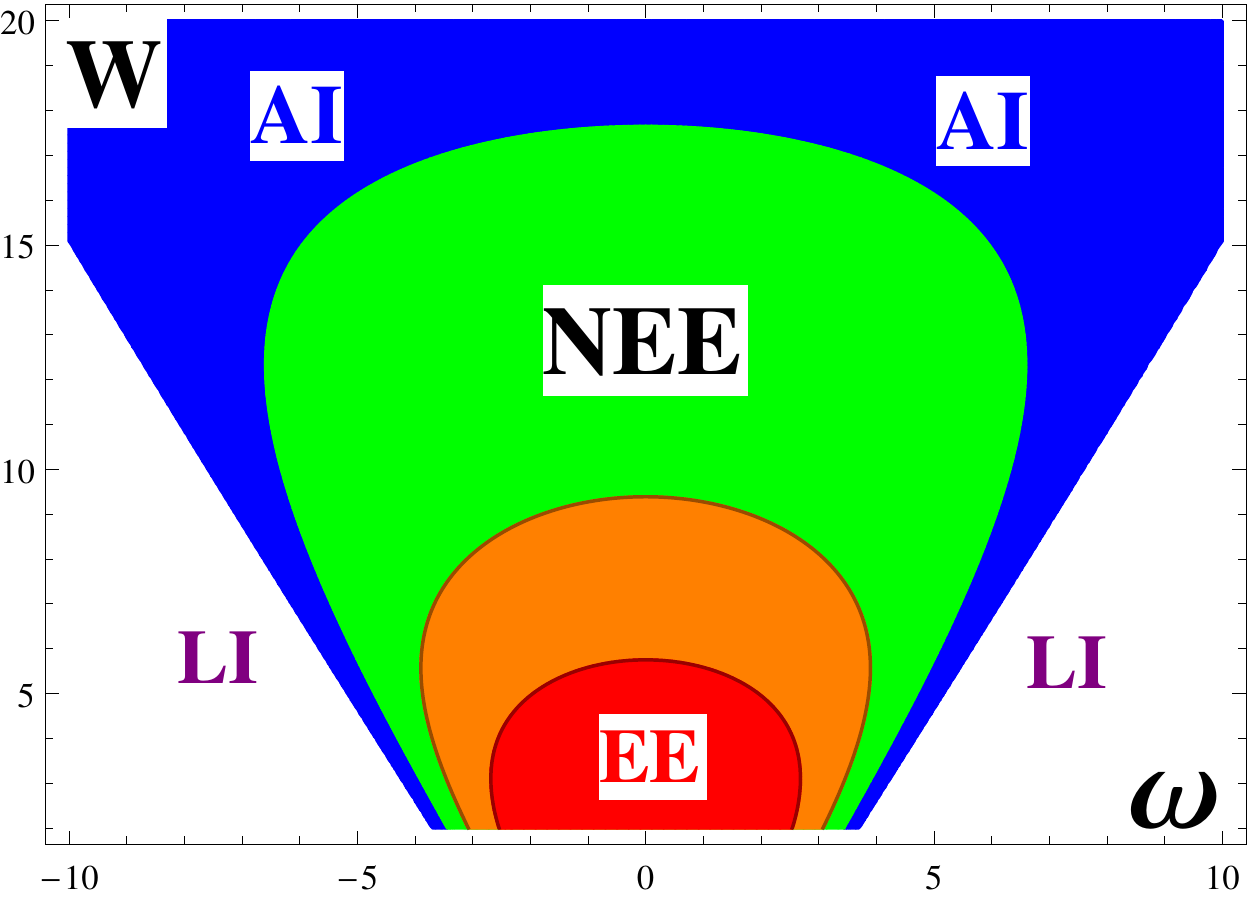}
  \includegraphics[width=0.45\linewidth]{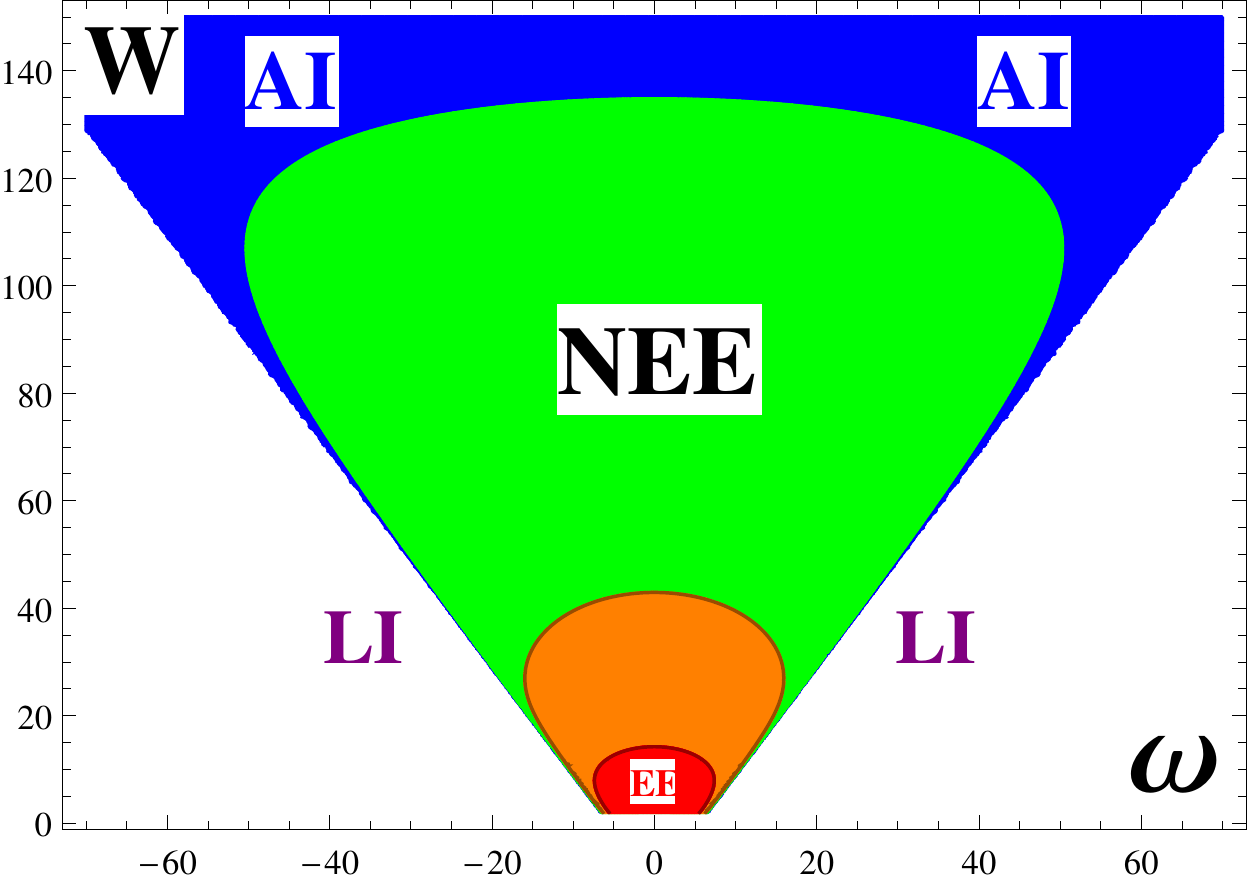}
 }
\caption{(Color online)   Sketch of the phase diagram in the $\omega-W$ plain for disordered BL.
Left panel: $K=2$. Right panel: $K=8$. The green area corresponds to the non-ergodic (multifractal) extended phase (NEE) where $1>D=D_{1}>0$ and $1>m_{0}>1/2$; the orange and  red area corresponds to the ergodic extended phase (EE) with $D=D_{1}=1$ and $m=1$; the blue area is the Anderson insulator (AI), corresponding to $D<0$, $D_{1}=0$ and $0<m_{1}<1/2$; the white area is the region where there are no eigenstates at $N=\infty$ but at a finite $N$ this is the region of Lifshitz insulator (LI) where $D=-\infty$, $D_{1}=0$ and $m_{1}=0$. The limits  of stability of the  Anderson insulator (upper arc) corresponds to    $\lambda=-\ln\langle|G| \rangle =\ln K$ and that of the non-ergodic extended phase (lower arc) corresponds to $\lambda_{typ}=-\langle \ln|G| \rangle =\frac{1}{2}\ln K$. The middle arc corresponds to $\lambda_{typ}= \ln K$. According to Ref.\cite{Warzel} for $\lambda_{typ}<\ln K$ there are only extended states. We believe that this is likely to be the line of the first order phase transition from NEE to EE phase. The top of the middle arc scales as $\sim K$ at large $K$, while that of the upper and lower arcs scale as $\sim K\,\ln K$ and $\sqrt{K}$, respectively. In our approximation (which is not accurate at small $W$) there is a mobility edge and a region of AI at  $W>2$ at $K=2$, while at $K=8$ and $W<50$ the  extended phase spreads up to  the spectral edge \cite{Warzel}.
}
 \label{Fig:phase-diagram}
\end{figure*}
%

One can apply Eqs.(\ref{Lambda-averaging}) to extend the above RSB solution to the case $\omega\neq 0$ and to obtain the phase diagram in the $\omega-W $ plain. For this purpose we employ the ansatz for $F(\epsilon)$ which generalizes Eq.(\ref{low-cutoff}) to the case of $\omega\neq 0$:
\begin{widetext}
\be\label{F-eps-omega}
F_{\rm eff}(\epsilon;\omega,W)= C(\omega,W)\,\left[\theta\left(\frac{W}{2} -
\epsilon\right) \theta\left(\epsilon-\omega-\frac{1}{\frac{W}{2}-\omega}\right) +
\theta\left(\frac{W}{2}+ \epsilon\right) \theta\left(\omega-\epsilon-\frac{1}{\frac{W}{2}+\omega}\right) \right],
\ee
where
$C(\omega,W)$ is the normalization constant.
\end{widetext}
Eq.(\ref{F-eps-omega}) is a good approximation only for $W/2 - |\omega| \gg \sqrt{K}$, and it is only qualitatively valid close to the edge of the spectrum.
However, it gives correctly the main characteristic features of the phase diagram.

More importantly, Eq.(\ref{F-eps-omega}) obeys the symmetry relation, Eq.(\ref{symmetry-omega}).
It appears that certain exact statements about the phase diagram  can be done just from the symmetry Eq.(\ref{symmetry-omega}) 
without using a concrete form of $F_{\rm eff}(\epsilon;\omega,W)$.

  As has been recently shown in Ref.\cite{Warzel}, it is convenient to describe the phase diagram   in terms of the Lyapunov exponent
 that characterizes the spacial structure of eigenfunctions. It is defined as:
\be\label{L}
\lambda_{typ}=-\lim_{r\to\infty}r^{-1}\left\langle \ln\left|\frac{\psi^{(g)}}{\psi^{(g-r)}}\right|\right\rangle,
\ee
where $r$ is the number of generations between the initial and the final point. Comparing the Schroedinger equation:
\be\label{Schroe}
\psi^{(g-1)}_{k(i)}+\sum_{j(i)=1}^{K}\psi^{(g+1)}_{j}=(\omega-\varepsilon_{i})\,\psi^{(g)}_{i}
\ee
with Eq.(\ref{eq:G_i^g(omega)}) one finds that:
\be\label{G-psi}
  G^{(g)}_{i} =\frac{\psi_{i}^{(g)}}{\psi_{k(i)}^{(g-1)}}.
\ee
As the result, the Lyapunov exponent is expressed through the Green's functions as follows:
\be\label{L-G}
\lambda=-\lim_{r\to\infty}r^{-1} \left\langle \sum_{k=g-r+1}^{g} \ln|G^{(k)}_{i_{k}}|\right\rangle,
\ee
or in terms of the effective PDF $F_{\rm eff}(\epsilon;\omega,W)$ of $G^{-1}=\omega -\epsilon$:
\be\label{L-eps}
\lambda_{typ}=\int F_{\rm eff}(\epsilon;\omega,W)\,   \ln|\epsilon-\omega|\,d\epsilon =-\langle \ln|G| \rangle.
\ee
We also define the log of the arithmetic average of $|G|$:
\be
\lambda=-\ln\langle |G| \rangle = -\ln\int F_{\rm eff}(\epsilon;\omega,W)\,\frac{d\epsilon}{|\omega-\epsilon|}.
\ee
It appears that the  limit of stability of Anderson insulator (AI) is naturally described through this quantity.
Indeed, we note that:
$$
\lambda=\ln K -\frac{1}{2}\Lambda\left(m=\frac{1}{2}\right).
$$
As in the case $\omega=0$, the optimal $m_{0}$ is equal to 1/2 on the entire line of the border of stability of AI
(the upper arc in Fig.\ref{Fig:phase-diagram}).
Since  the limit of stability of AI corresponds to   $\Lambda(m_{0})=0$ we obtain on the entire upper arc of Fig.\ref{Fig:phase-diagram}:
\be\label{lambda-up}
\lambda=\ln K.
\ee

Now consider the limit of stability of the non-ergodic extended phase (lower arc on Fig.\ref{Fig:phase-diagram}). It corresponds to $m_{0}=1$, i.e.
\be\label{low-arc-der}
 \partial_{m}\Lambda(m)|_{m=1}=0.
\ee
As for the case $\omega=0$, one can show using the symmetry Eq.(\ref{symmetry-omega}) that $\Lambda(m=1)=\ln K$. Then with the help of the same 
symmetry we obtain from Eqs.(\ref{Lambda-averaging}),(\ref{low-arc-der}):
\bea\label{lambda-typ}
&&\partial_{m}\Lambda(m)|_{m=1}=-\Lambda-2\int F_{\rm eff}(\epsilon;\omega,W)\,\frac{\ln|\omega-\epsilon|}{|\omega-\epsilon|^{2}}\, d \epsilon =
\nonumber \\ &=&-\ln K + 2\int F_{\rm eff}(\epsilon;\omega,W)\,\ln|\omega-\epsilon|\,d\epsilon = 0.
\eea
This implies that
at the limit of stability of the non-ergodic extended phase (lower arc in Fig.\ref{Fig:phase-diagram}) the Lyapunov exponent is equal to:
\be\label{lambda-low-arc}
\lambda_{typ}=\frac{1}{2}\ln K.
\ee
This result is expected, as the  Lyapunov exponent  in the ergodic phase on BL is not zero, which is a consequence of the exponential growth of the number of sites with $r$. Indeed, the Green's function for the Laplace operator \cite{Chung-Yau}  on BL at $r\gg 1$ reads:
\be\label{Laplace}
G(r;\omega)=-i\frac{K^{-r/2}}{K-1}{\rm exp}(i r\kappa),\;\;\;\;\kappa=\sqrt{\frac{2\sqrt{K}-|\omega|}{\sqrt{K}}}.
\ee
Therefore inside the energy band $|\omega|\leq 2\sqrt{K}$ of the pristine infinite BL the   Lyapunov exponent is:
\be\label{L-clean}
\lambda_{{\rm erg}}=\frac{1}{2}\ln K,
\ee
which is a minimal Lyapunov exponent on BL.

The lower bound of the Lyapunov exponent,  Eq.(\ref{lambda-low-arc}), was earlier found  in Ref. \cite{Warzel}. However, its physical meaning remained somewhat unclear. We now claim that this is the limit of stability of NEE phase with respect to NEE-EE transition.

There is a rigorous mathematical result \cite{Warzel} that inside the spectrum $|\omega| < 2\sqrt{K}+W/2$ of an infinite tree the imaginary part of the Green's function is finite at $\eta\rightarrow 0$ provided that
\be\label{middle-arc}
\lambda_{typ}<\ln K.
\ee
This implies  existence of only ergodic wave functions under the condition Eq.(\ref{middle-arc}). The equality in Eq.(\ref{middle-arc}) corresponds to the middle arc in Fig.\ref{Fig:phase-diagram}. For $K\geq 2$ it lies inside of the region of stability of both RSB and RS solutions. Therefore it is natural to interpret the middle arc as the line of the first order phase transition from NEE to EE phase. In the middle of the band $\omega=0$ our approximation Eq.(\ref{low-cutoff}) gives the following values of $W_{E}$ for the first order ergodic transition point:
\begin{table}[h!]
  \centering
  \caption{ Values for $W_{E}$ at different $K$ in the center of the band}
  \label{tab:table1}
  \begin{tabular}{|c||c|c|c|c|c|c|c|}
   \hline
    K & 2 & 3 & 4 & 5 & 6 & 7 & 8\\
    \hline
     $W_{E}$  & 9.4 & 15.2& 20.8 & 26.4 & 31.9 & 37.4 & 42.9 \\
    \hline
\end{tabular}
\end{table}
\\Note a rather good coincidence with $W_{E}=9.9$ at $K=2$ reported in Ref.\cite{Arx}.

Concluding this Section we would like to emphasize the difference between the phase diagram for $K=2$ and $K=8$ in Fig.\ref{Fig:phase-diagram}. One can show that the limit of stability of NEE phase with respect to NEE-EE transition $W_{0}\propto \sqrt{K}$, while the AT point $W_{c}$ scales like $K\ln K$ and $W_{E}\propto K$. That is why the area of the EE phase relative to that of NEE phase shrinks   as $K$ increases.

This can be interpreted as relative insignificance of the EE phase in the classical limit. Indeed, the parameter that quantifies the quantum-to-classical crossover is the ratio of the typical potential energy, the on-site energy fluctuations, to the typical kinetic energy, the bandwidth. In our case it is equal to $r_{s}=W/\sqrt{K}$. The AT point corresponds to
$r_{s}=W_{c}/\sqrt{K} \sim \sqrt{K}\ln K$, so that the classical limit is the limit of large $K$. Our results show that in this limit the NEE phase, which in many respects is similar to the glass, is occupying the lion share of the phase diagram, while the AI phase is gone to very large $W$.

\section{Anderson and Lifshitz insulators}
\begin{figure}[h!]
\center{
\includegraphics[width=0.9\linewidth]{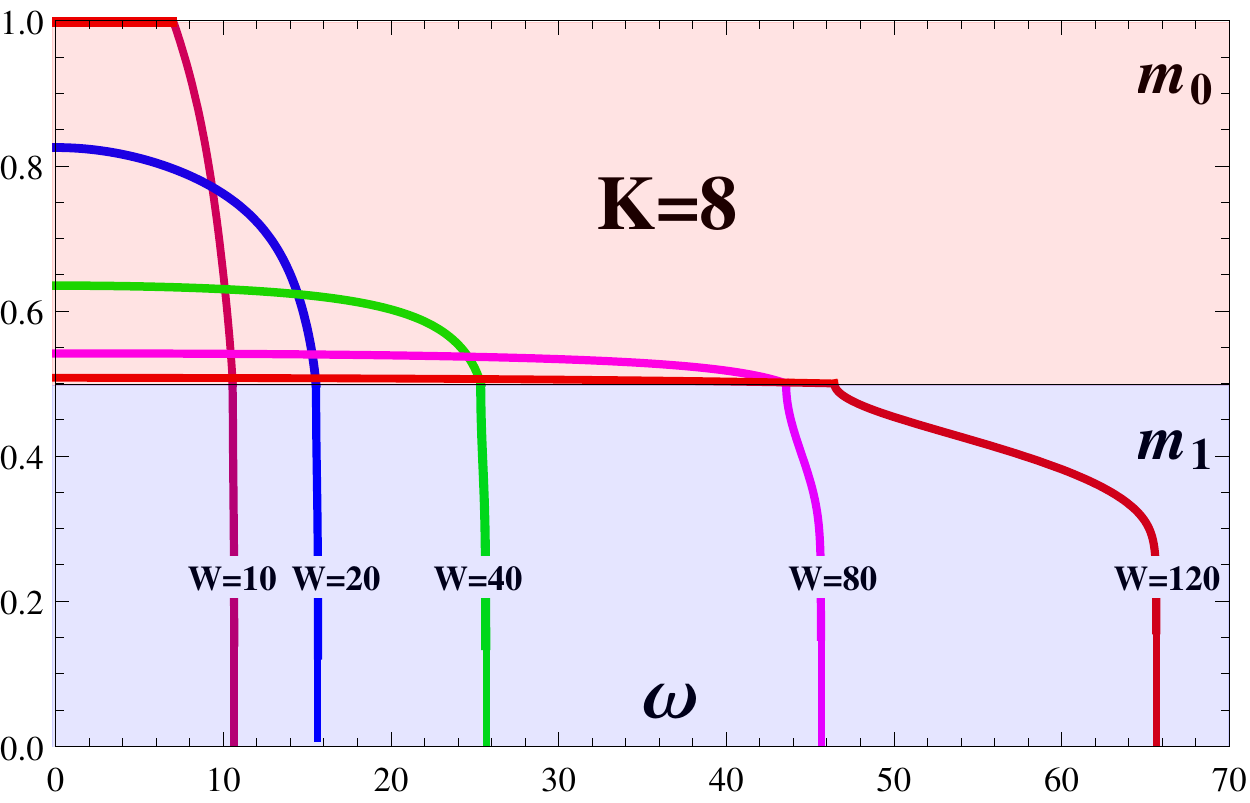}
}
\caption{(Color online) The solution $m_{1}$   to the equation $\Lambda(m_{1})=0$ in the region of localized  states (blue) and the solution $m_{0}$ to the equation $\partial_{n}\Lambda(m_{0})=0$ in the region of extended states (rose) as  functions of energy $\omega$ at fixed disorder $W=10,20,40,80,120$ for $K=8$. For $W=10, 20, 40$ the drop of $m_{1}$  from $m_{1}=1/2$ to $m_{1}=0$ is practically vertical in our approximation Eqs.(\ref{Lambda-averaging}),(\ref{F-eps-omega}). Thus at $W<50$ the transition from  the extended (NEE or EE) phase with $1/2<m_{0}\leq 1$ to the Lifshitz insulating phase (LI) phase with $m_{1}=0$ happens directly \cite{Warzel} avoiding the intermediate Anderson insulating phase (AI) with $0<m_{1}<1/2$.
}
\label{Fig:m1-omega}
\end{figure}
As was shown in Appendix A, the solution $m_{1}$ of the equation $\Lambda(m_{1})=0$ determines the power-law distribution $\tilde{P}(x=N|\psi|^{2})\propto x^{-(1+m_{1})}$ of the wave function amplitudes in the insulating phase. For the Anderson insulator in a finite-dimensional space this power is -1.  The reason is that
the wave function amplitude  drops down exponentially with the distance $r$ from the localization center and the number of sites increases only as a power-law. There are two cases where a non-trivial power may arise: (a) power-law localization in finite-dimensional systems with long-range hopping (Levy flights) \cite{Shlyapnikov}
and (b) exponential localization on BL \cite{Our-BL}. In both cases this non-trivial power-law is the result of competition between the $r$-dependence of the volume and that of the wave function amplitude. On  BL this makes it possible for two different insulating phases to exist. One of them is an Anderson insulator with a non-trivial  $1/2 > m_{1} >0$, and the other one is the {\it Lifshitz insulator} in the region of the Lifshitz tail at the edge of the spectrum. The latter phase corresponds to $m_{1}=0$, since the number of accessible sites does not grow exponentially any more due to extremely small probability of resonances in the Lifshitz tail region.

There is a rigorous mathematical result \cite{Warzel} that there exists a minimal disorder strength $W_{\rm min}\sim K$ such that for $W<W_{\rm min}$ the Anderson insulator does not exist at all, and extended states survive up to the spectral edge. In our language this would mean that $m_{1}$ jumps discontinuously at the spectral edge from  $m_{1}\geq 1/2$  to $m_{1}=0$ as the energy $\omega$ increases, i.e. a direct transition from the extended to Lifshitz insulator phase occurs.

Although our approximation for $F(\epsilon;\omega,W)$ Eq.(\ref{F-eps-omega}) does not provide  quantitatively accurate results near the onset of the Lifshitz tail (at $|\omega|\approx 2\sqrt{K}+W/2$), qualitatively the behavior obtained in our RSB scheme  is similar to the one predicted in Ref.\cite{Warzel}. In Fig.\ref{Fig:m1-omega} we present the results of RSB calculations, Eqs.(\ref{Lambda-averaging}),(\ref{F-eps-omega}) for $K=8$, for $m_{1}(\omega)$ at fixed disorder $W$. One can see that at small enough $W$ there is practically a jump from $m_{1}=1/2$ at the onset of an insulating phase to $m_{1}=0$. Only for large enough disorder ($W>\sim 60$ for $K=8$) the region of smooth variations of $m_{1}$ between $M_{1}=1/2$  and $m_{1}=0$ (which corresponds to the Anderson insulating phase) becomes visible (see also Fig.\ref{Fig:phase-diagram}).

\section{Discussion}
Very recently during our work on the manuscript two preprints Refs.\cite{TihMir, Lemarie} appear in which the authors claim that 
the case of the finite Cayley tree is different from the case of a finite RRG. While in the first case they agree that the non-ergodic 
multifractal phase does exist, they insist that in the case of RRG only ergodic phase is present at any $W<W_{c}$ and $N\rightarrow \infty$. 
In this connection we would like to recall that the derivation of Eq.(\ref{Lambda-D}) explicitly uses an {\it assumption} that the exponential 
growth in Eq.(\ref{Lambda}) is terminated as soon as the first loop is completed on RRG, i.e. $\ell=\ell_{t}$ reaches the diameter of the graph $R$. 
Should the process of  growth of $\Im G$ continue further at $\ell >R$, the spectral fractal dimension $D$ may be larger than that given by 
Eq.(\ref{Lambda-D}) and may in principle reach the value of 1. An argument against this scenario is that Eq.(\ref{Lambda-D}) works perfectly 
well for the case of the Rosenzweig-Porter random matrix model.  In this case it corresponds to the termination of the exponential growth 
just after one iteration ($\ell_{t}=1$), as the next one would for sure lead to a completed loop. We do not see a reason why for the RP 
model the first completed loop should lead to the termination while for RRG it should not.

One of the comments concerning Ref.\cite{Lemarie} is that at a small (positive) $K-1$, the order of arcs in Fig.\ref{Fig:phase-diagram} 
may be reversed (and in fact it is reversed in our crude approximation), so that the "middle arc" of the first order phase transition becomes 
larger than the "upper arc" of the limit of stability of AI. In this case the NEE phase is not realized at all, and the first order transition 
happens from AI to EE phase with the jump of $D_{q}$ from 0 to 1 at $W=W_{E}$, as was reported in Ref.\cite{Lemarie} for a model with the average 
connectivity $K=1.12$.

As for the old analytical theory \cite{MF-ergodic-theory} which predicts only an ergodic phase for $W<W_{c}$, we 
believe that the situation is very similar to the one which concerns the problem of a classical spin glass on a BL. 
There was a long discussion on this matter summarized very well in a paper by Mezard and Parisi \cite{Mezard-Parisi}. 
In the problem of spin glass there is the Bethe-Peierls "solution" which is similar in spirit to the theory of Refs. \cite{MF-ergodic-theory}. 
However, this solution appears to be  wrong (as it fails to identify the magnetic field-driven transition) for the RRG and is valid only for 
restricted number of models on a finite tree with very special boundary conditions. One can show \cite{Mezard-Parisi} that this solution, like 
the solution of Refs.\cite{MF-ergodic-theory}, is identical to replica symmetric solution in the formulation using the replica trick. One can 
also show that the solution with the broken replica symmetry (even on a one-step RSB level) is a correct one for the spin glass problem on RRG. 
We believe that this paradoxical situation when the "mathematically clean" Bethe-Peierls solution is not appropriate to spin-glass physics and a 
full of dangerous tricks RSB solution is physically correct, is repeating now in the field of Anderson (and many-body) localization. 

The key point of our analysis is the proof of {\it co-existence} of both RS solution that describes the ergodic (EE) phase  and the RSB solution 
that describes the non-ergodic extended (NEE) (multifractal) phase. A competition between these two solutions is the main issue of the problem. 
Based on recent rigorous mathematical results \cite{Warzel} we made a conjecture that the line of the first order transitions between   NEE and EE phases 
  corresponds to the Lyapunov exponent $\lambda_{typ}=\ln K$. For $K=2$ this 
assumption agrees well with $W_{E}$ found in Ref.\cite{Arx} by numerical diagonalization of large RRG.

\section{Conclusion}
In this paper we derive an expression, Eq.(\ref{Lambda-D}), for the spectral dimension $D$ which governs the typical 
imaginary part $\Im G$ of the  Green's functions on the Bethe lattice.  
  We prove existence of the extended non-ergodic phase on BL  
and show that the spectral fractal dimension in this phase coincides with the fractal dimension $D_{1}$ of the 
wave functions support set. We also prove that Eq.(\ref{Lambda-D}) applies not only to BL but also to random 
systems with infinite connectivity (e.g. to the  Rosenzweig-Porter random matrix theory). The unifying concept for all these models 
is  the self-consistent theory of Abou-Chakra, Thouless and Anderson \cite{AbouChacAnd}.

We develop a replica approach with a one-step replica symmetry breaking which 
allows us to suggest an approximation for critical disorder for the Anderson model on BL with 
an arbitrary branching number $K\geq 2$ and the box probability distribution of the random on-site energies. 
This approximation appeared to be the best available so far. It allows us also to obtain a  phase diagram for 
the localization on BL which obeys certain constraints  proven rigorously in mathematical literature \cite{Warzel}. 
In particular we uncover the physical meaning of the line on the phase diagram where the typical Lyapunov exponent is 
equal to $\frac{1}{2}\,\ln K$ \cite{Warzel} as the limit of stability of the non-ergodic extended phase with respect to transition to the ergodic phase.
We also conjecture that the first order phase transition between these phases happens prior to this limit is reached when the typical 
Lyapunov exponent is equal to $\ln K$.

Finally, we suggest existence of two types of insulators on  a finite BL, the Anderson and the Lifshitz insulators, 
and characterize them unambiguously in terms of the parameter $m$ of the replica symmetry breaking.  
  For a infinite BL with large branching number and weak disorder we confirm survival  of the extended phases   \cite{Warzel} all the way up to
 the spectral edge.

\section{Acknowledgement}
We appreciate discussions with G. Biroli, J. T. Chalker, E. Cuevas, M. Feigelman, A. Yu. Kitaev,  G. Parisi,  M. Tarzia, K. Tikhonov. 
We are especially grateful to E. Bogomolny and S. Warzel who helped us to understand importance and pertinence of the recent rigorous 
mathematical results on localization on BL. A support from LPTMS of University of Paris-Sud at Orsay (V. E. K.), 
College de France and ICTP (Trieste) (B. L. A.)  where important part of this work was done, is gratefully appreciated. The research of 
L.B.I. was partially supported  by the Russian Science
Foundation grant No. 14-42-00044.

\appendix
\numberwithin{equation}{section}
\section{}
The probability distribution function $P(\bar{\rho})$ for $W<W_{c}$ and $\eta<\eta_{c}$ coincides (in the leading in $\ln N\gg 1$ approximation) with that of the normalized amplitude of the wave function $x=N|\psi_{a}|^{2}$ averaged over the energy interval $\Delta E_{a}\sim  \eta_{c}$. Indeed, in the entire extended phase $\Delta E_{a}\sim N^{-z}\gg \delta$, as the dynamical exponent $z<1$. Then $\bar{\rho}=\rho/\langle \rho\rangle$ is dominated by eigenstates with $|\omega-E_{a}|<\eta_{c}$ and is given (up to a prefactor of order 1) by:
\bea\label{x-rho}
\bar{\rho}&\sim &\langle \rho\rangle^{-1}\,\eta_{c}^{-1}\sum_{a, |\omega-E_{a}|<\eta_{c}}|\psi_{a}(i)|^{2}=\nonumber \\
&=& \frac{1}{\langle \rho\rangle\,\delta}\,\overline{|\psi|^{2}}=N\,\overline{|\psi|^{2}},
\eea
where $\overline{|\psi|^{2}}$ denotes the wave function amplitude averaged over an interval of energies $|\omega-E_{a}|\sim \eta_{c}\gg \delta$.
Eq.(\ref{x-rho}) ensures that the functions $f(\alpha)$ in the multifractal ansatz Eq.(\ref{multifractal-ansatz}) for $\bar{\rho}$ and $x=N\,\overline{|\psi|^{2}}$ are identical in the entire region of extended states.
\begin{figure}[h]
\center{
\includegraphics[width=0.9\linewidth]{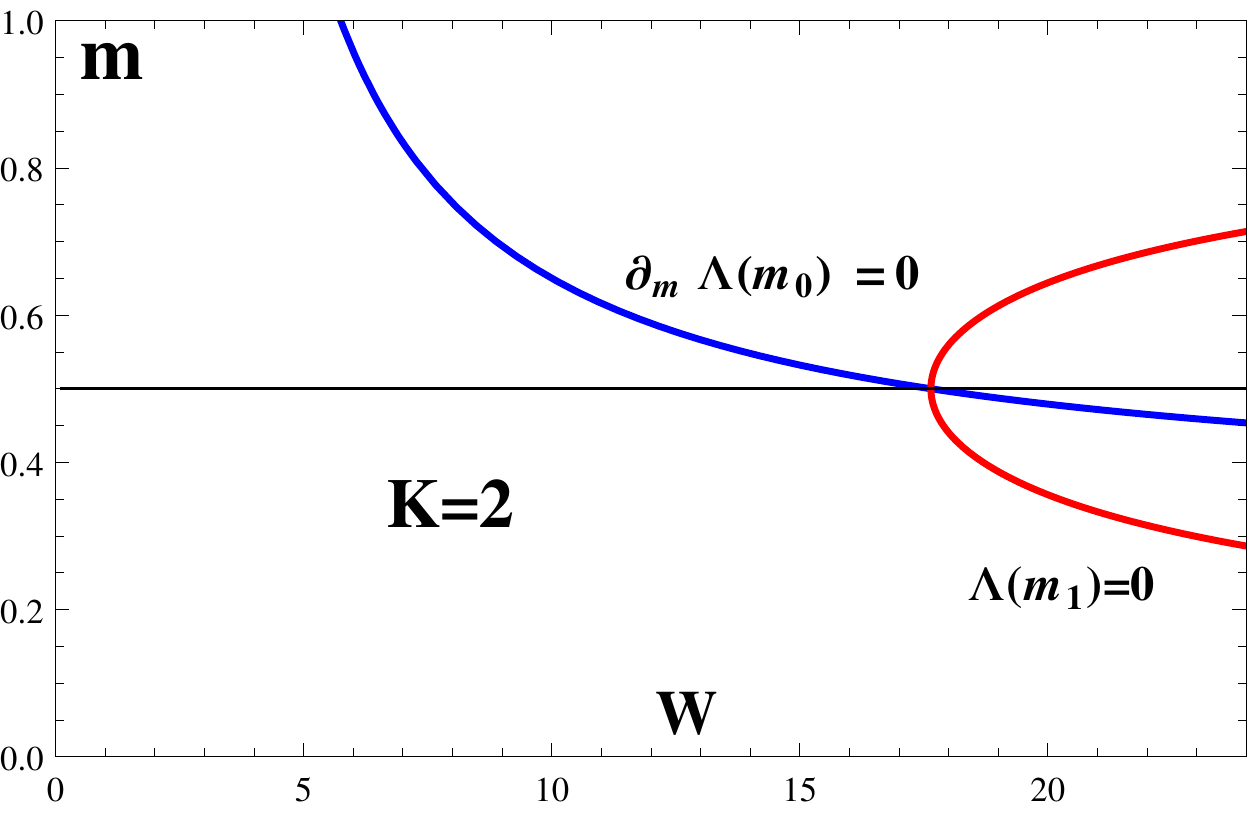}
}
\caption{(Color online) The solutions to the equations $\partial_{m}\Lambda(m_{0})=0$ (blue) and $\Lambda(m_{1})=0$ (red) 
for $\omega=0$ using $F_{\rm eff}(\epsilon)$ from Eq.(\ref{low-cutoff}). At the AT point $W=W_{c}$ they all coincide $m_{0}=m_{1}=1/2$. 
For $W>W_{c}$ the lower branch of the red curve describe the slope of $f(\alpha)$ for $\alpha=1-\ln(N\,|\psi|^{2})/\ln N$. 
The termination point of the RSB solution corresponds to $m_{0}=1$.
}
\label{Fig:m0-m1}
\end{figure}

Note that  such an averaging over many states eliminates the fast oscillations at a scale of the de Broglie wavelength and is equivalent to the procedure of extraction of the "envelope" of $|\psi_{a}|^{2}$ suggested in  Ref.\cite{Our-BL}. It is only meaningful if (a) $\eta_{c}\gg \delta$ and (b) the envelopes of different wave functions are strongly correlated in the energy interval $\sim \eta_{c}$. The latter property is well known for the multifractal (extended, non-ergodic) states \cite{Chalker,KrMut}.

The above arguments fail in the localized phase $W>W_{c}$. In this phase, the PDF of $\bar{\rho}$ and $N\,\psi^{2}$ can still obey the multifractal ansatz
\cite{Our-BL} but the corresponding $f(\alpha)$ are different. The symmetry Eq.(\ref{MF-sym}) is still valid for $f(\alpha)$ of $\bar{\rho}$ but is violated for $f(\alpha)$ describing fluctuations of $|\psi|^{2}$.

Note that the latter was studied in Ref.\cite{Our-BL} in the "directed polymer" approximation valid at large $W\gg 1$, and the result was that $f(\alpha)=k\,\alpha$ is linear with the termination point $\alpha_{>}=k^{-1}$. In this approximation, the slope $k$ can be  found from the equation
\be
\ln\left(\frac{W}{2} \right)=\frac{1}{2k}\,\ln\left(\frac{K}{1-2k} \right),
\ee
which coincides with Eq.(\ref{eq:lnWc/2}) at $W=W_{c}$ and with
\be\label{m1}
\Lambda(m_{1})=0,\;\;\;\;m_{1}=k,
\ee
for $W>W_{c}$. On the other hand, the linear $f(\alpha)$ with the slope $k$, implies a power-law PDF:
\be
\tilde{P}(x)\propto x^{-(1+k)}.
\ee
We conclude that the solution $m_{1}$ to Eq.(\ref{m1}) determines the power $-(1+m_{1})$ in the power-law PDF of $x=N\,|\psi|^{2}$ at $W>W_{c}$.

Note that real solutions to Eq.(\ref{m1}) at $\omega=0$ exist only for $W>W_{c}$ (see Fig.\ref{Fig:m0-m1}) and it is the smaller of them  that determines the slope $k$ \cite{Our-BL}.
\section{}
In this Appendix  we will show where the symmetry Eqs.(\ref{low-cutoff}),(\ref{symmetry-omega}) comes from and describe an alternative way of 
deriving Eq.(\ref{Lambda-averaging}) in which the symmetry is automatically present.

Let us cast Eq.(\ref{eq:G_i^g(omega)}) as
\begin{equation}\label{G-eq}
G^{-1}_{k-1}(i_{k-1}) = \omega -E_{i_{k-1}}
-G_{k}(i_{k}),
\end{equation}
where $G_{k}(i_{k})$ is the Green's function in a point $i_{k}$ of the $k-th$ generation, and we introduced the notation:
\be\label{E}
E_{i_{k-1}}=\varepsilon_{i_{k-1}}+\sum_{j(i_{k-1}), j\neq i_{k}} G_{k}(j).
\ee
\begin{figure}[h]
\center{
\includegraphics[width=1.1\linewidth]{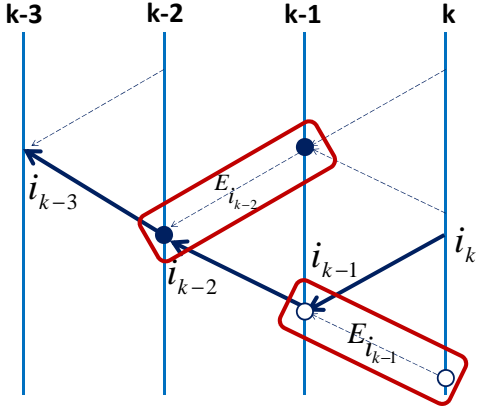}
}
\caption{(Color online) Sites involved in $E_{i_{k-1}}$ (open circles) and $E_{i_{k-2}}$ (full circles) belong to different branches of the tree and thus $E_{i_{k-1}}$ and $E_{i_{k-2}}$ are statistically independent. Vertical lines are generations, the fat solid arrows denote the path, the dashed arrows denote links other than those belonging to the path.
}
\label{Fig:path}
\end{figure}
The reason to introduce the set of $E_{i_{k}}$ is that these quantities are statistically independent at different sites of the path (see Fig.\ref{Fig:path}).  Then the measure $d\mu=\prod_{k=1}^{\ell} dE_{i_{k}}F_{0}(E_{i_{k}})$ along a path is:
\be\label{measure}
 d\mu=\prod_{k=1}^{\ell} dx_{k}\,F_{0}\left(\omega-x_{k}-\frac{1}{x_{k+1}}\right),
\ee
where $x_{k}=G^{-1}_{k}(i_{k})$ and $F_{0}(E_{i_{k}})$ is the PDF of $E_{i_{k}}$ which is independent of $i_{k}$ in the bulk of an infinite tree.

Consider the probability distribution function $P(X)$ of
$$
X=\frac{1}{\ell}\sum_{k=1}^{\ell}\ln|G_{k}(i_{k})|.
$$
where the sum is running along some path. Then the probability distribution ${\cal P}(y)$ of $y=\prod_{k=1}^{\ell} |G_{k}(i_{k})|^{-1}$ is:
\be
{\cal P}(y)=(1/\ell)\,P(X)\,e^{\ell X}\left|_{X=\ell^{-1}\,\ln(1/y)}\right.
\ee

The  function $P(X)e^{\ell X}$ can be represented as:
\begin{widetext}
\bea
P(X)e^{\ell X}&=&\int \prod_{k=1}^{\ell} \frac{dx_{k}}{|x_{k}|}\,F_{0}\left(\omega-x_{k}-\frac{1}{x_{k+1}}\right)\,\delta\left(X+\ell^{-1}\sum_{m=1}^{\ell}\ln|x_{m}|\right)\label{P-X}\\
&=& \int_{-\infty}^{+\infty}\frac{d p}{2\pi}\,e^{i p X}\,\int \prod_{k=1}^{\ell}\frac{dx_{k}}{|x_{k}|}\, e^{i p \ell^{-1} \ln|x_{k}|}\;F_{0}\left(\omega-x_{k}-\frac{1}{x_{k+1}}\right)\label{P-X-p}.
\eea
\end{widetext}
Making the transformation
$$
X\rightarrow -X,\, p\rightarrow -p,\, x_{k}\rightarrow\frac{1}{x_{k}}
$$
one obtains:
\begin{widetext}
\be\label{rev-P-X}
P(-X)e^{-\ell X}
= \int_{-\infty}^{+\infty}\frac{d p}{2\pi}\,e^{i p X}\,\int \prod_{k=1}^{\ell}\frac{dx_{k}}{|x_{k}|}\, e^{i p \ell^{-1} \ln|x_{k}|}\;F_{0}\left(\omega-x_{k+1}-\frac{1}{x_{k}}\right).
\ee
\end{widetext}
The only difference between Eq.(\ref{P-X-p}) and (\ref{rev-P-X}) is the reversed order of $x_{k}$ and $1/x_{k+1}$ in the arguments of $F_{0}$.
The last step is to make re-ordering the variables $x_{k}$ as follows:
$$
x_{1}\rightarrow x_{\ell},\;\;x_{2}\rightarrow x_{\ell-1},  ...\;\; x_{\ell-1}\rightarrow x_{2},\;\;x_{\ell}\rightarrow x_{1}
$$
Then Eqs.(\ref{P-X-p}),(\ref{rev-P-X}) become identical which completes the proof of the symmetry:
\be\label{sym}
P(X) e^{\ell X} = P(-X)e^{-\ell X},\;\;\Leftrightarrow \;\;{\cal P}(y)={\cal P}(1/y).
\ee
Using Eq.(\ref{all-pathes}) we can express $\Lambda(\omega,m)$ in terms of ${\cal P}(y)$, instead of Eq.(\ref{Lambda-averaging}):
\begin{widetext}
\be\label{RSB-eq}
\Lambda(\omega,m)=\frac{1}{m}\ln\left[K \left\langle \prod_{k=1}^{\ell}|G_{k}(i_{k})|^{2m}\right\rangle^{\frac{1}{\ell}} \right]=\frac{1}{m}\ln\left\{K\left[\int \frac{dy}{y^{2m}}\,{\cal P}(y)\right]^{\frac{1}{\ell}}\right\}.
\ee
\end{widetext}
At large $\ell\rightarrow \infty$ it is convenient to introduce the function $p(z)$ such that:
\be\label{p}
[p(z)]^{\ell}={\cal P}(z^{\ell}),\;\;\;\;\;p(z)=p(1/z),
\ee
and evaluate the integral in Eq.(\ref{RSB-eq}) in the saddle-point approximation. Then we obtain in the limit $\ell\rightarrow\infty$:
\be\label{saddle}
\left[\int \frac{dy}{y^{2m}}\,{\cal P}(y)\right]^{\frac{1}{\ell}} = p(z_{m})\,z_{m}^{1-2m},
\ee
where the saddle-point $z_{m}$ is the solution to:
\be\label{stat}
\partial_{z}\ln p(z)+\frac{(1-2m)}{z}=0.
\ee
The normalization of $\int {\cal P}(y)\,dy =1$ then imposes the normalization of $p(z)$:
\be\label{norm-p}
p(z_{0})\,z_{0}=1.
\ee
Comparing Eqs.(\ref{RSB-eq}),(\ref{p}) with Eq.(\ref{Lambda-averaging}) one concludes that:
\be\label{correspon}
\tilde{I}_{m}=\int F_{\rm eff}(\epsilon+\omega)\,\frac{d\epsilon}{|\epsilon|^{2m}} \Rightarrow I_{m}=p(z_{m})\,z_{m}^{1-2m}.
\ee
Note that in the present --more regular-- derivation, the knowledge of   ${\cal P}(y=\prod_{k=1}^{\ell} |G_{k}(i_{k})|^{-1})$ (and thus of $p(z)$) 
is sufficient to find the analogue of the integral  $I_{m}$ in Eq.(\ref{Lambda-averaging}) without any reference to $F_{\rm eff}(\epsilon)$. Moreover, the 
symmetry $p(z)=p(1/z)$ proven above is sufficient to prove the symmetry:
\be\label{m-symm}
z_{m}=z_{1-m}, \Rightarrow I_{m}=I_{1-m},
\ee
without using the symmetry $F_{\rm eff}(E)=F_{\rm eff}(1/E)$.

It follows immediately from Eq.(\ref{m-symm}) that $I_{1}=I_{0}=1$, which helps to prove that $D=1$ at $m=1$, i.e. the existence of the replica-symmetric 
solution. It is also sufficient to prove Eq.(\ref{lambda-up}). Another useful relation which follows from Eq.(\ref{m-symm}) is:
\be\label{m-der-id}
 \partial _{m}I_{m} = - \partial_{m} I_{1-m}.
\ee
Eq.(\ref{m-der-id}) is sufficient to prove that $m=1/2$ at the AT point. It is also operative to prove Eq.(\ref{lambda-low-arc}).

\begin{figure}[h!]
\center{
\includegraphics[width=0.9\linewidth]{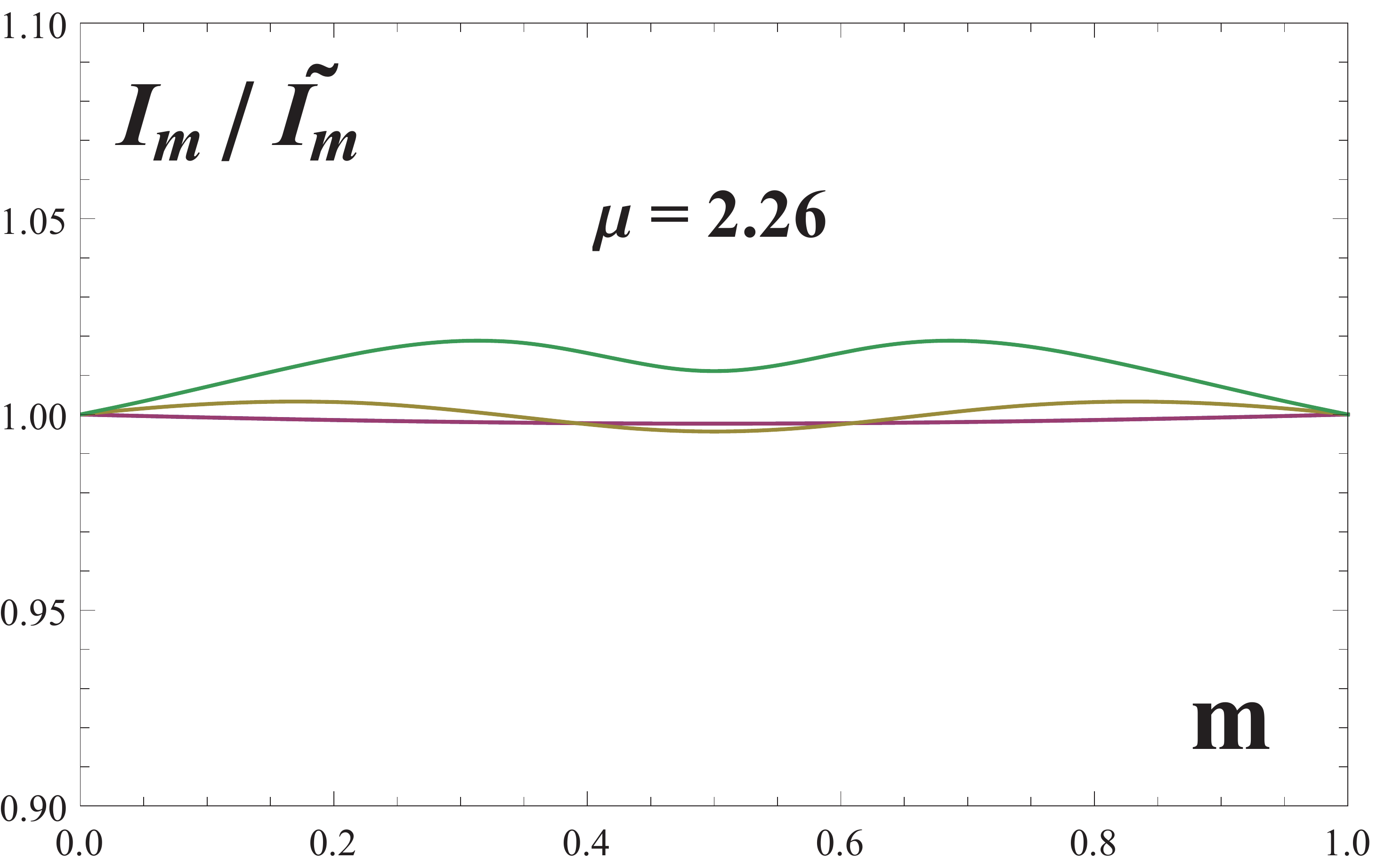}
}
\caption{(Color online) The ratio $I_{m}/\tilde{I}_{m}$ of $I_{m}=p(z_{m})z_{m}^{1-2m}$ with $p(z)$ from
 Eq.(\ref{p-approx}) and $\tilde{I}_{m}=\int F_{\rm eff}(\epsilon)\,|\epsilon|^{-2m}\,d\epsilon$ with $F_{\rm eff}(E)$ given by Eq.(\ref{low-cutoff}). 
Because of the symmetry of both $I_{m}$ and $\tilde{I}_{m}$ w.r.t. $W/2 \rightarrow 2/W$, only $W>1$ are plotted: $W=4, 40, 400$. 
Thus the difference between $I_{m}$ and $\tilde{I}_{m}$ does not exceed $2\%$ in an exponentially broad interval of $W$.
}
\label{Fig:approx}
\end{figure}

We conclude this Appendix by evaluating   $p(z)$ for the simplest  approximation  equivalent to Eq.(\ref{low-cutoff}). We start by
an even simpler task of computing ${\cal P}(y)$ and $p(z)$ neglecting the real part of the self-energy $\Re \Sigma$. In this case one substitutes
$F_{0}(\omega-x_{k}-x_{k+1}^{-1})$ for $F_{0}(\omega-x_{k})$ in Eq.(\ref{measure}), where $F_{0}(x)=W^{-1}\,\theta(W/2 -|x|)$. 
Then Eq.(\ref{P-X-p}) at $\omega=0$ results in:
 \be\label{calP-sigma-no}
{\cal P}(y)=\frac{\ell^{\ell -1}}{(\ell -1)!}\,\left( \frac{2}{W}\right)^{\ell}\,\left[ \ln\left( \frac{W}{2y^{\frac{1}{\ell}}}\right)\right]^{\ell -1},
\ee
and
\be\label{p-no-Re}
p(z)=\frac{2e}{W}\,\ln\left(\frac{W}{2z} \right).
\ee
Eq.(\ref{p-no-Re}) does not respect the symmetry  Eq.(\ref{p}), because $\Re \Sigma$ is neglected. The cheapest way to fix this drawback is
to look for an {\it approximate} solution of the form:
\be\label{p-approx}
p(z)= A\,\left[\ln\left(\frac{W}{2z} \right)\,\ln\left(\frac{W z}{2} \right)\right]^{\mu},
\ee
with some free parameter $\mu$ and the normalization constant $A$ found from Eq.(\ref{norm-p}).
It appears that the choice $\mu\approx 2.3$  results in $I_{m}$ (see Eq.(\ref{correspon})) which is an excellent approximation at any $0< m\leq 1$ 
and in the exponentially broad range of $W$ (see Fig.\ref{Fig:approx}) for the integral $\tilde{I}_{m}$ found from $F_{\rm eff}(\epsilon)$ of the 
form Eq.(\ref{low-cutoff}) for $W>2$ and analytically continued to $0<W<2$.

The results of this Appendix demonstrate that the notion of "effective distribution of on-site energies" $F_{\rm eff}(\epsilon)$ is a 
convenient presentation trick but it is not necessary to obtain all the results of the paper. They can be formulated entirely in terms 
of $I_{m}$ given by Eq.(\ref{correspon}) and obeying the symmetry Eq.(\ref{m-symm}). This symmetry is equivalent to the symmetry 
w.r.t. $\beta\rightarrow 1-\beta$, as formulated in Eq.(6.8) of Ref.\cite{ACA}.


\begin{thebibliography}{99}
\bibitem{almost-surely} This means that the probability to find a loop with the length $\ell <R$ is exponentially small and that for $\ell >R$ is exponentially close to 1.
\bibitem{accuracy} Writing $\ln W_{c} =\ln (K \ln K)\,[1 +  \ln C /\ln (K \ln K)]$, where $C$ in the pre-factor in Eq.(\ref{lKWc}) one concludes that the accuracy of the approximation is controlled by the small parameter
$1/\ln K$.
\end{thebibliography}
\end{document}